 \documentclass[prl,amsmath,amssymb,twocolumn, showpacs, superscriptaddress,10pt]{revtex4-1}

\usepackage{amsmath}
\usepackage{hyperref}
\usepackage{graphicx}
\usepackage{amsfonts}
\usepackage{amsthm}
\usepackage{cases}
\usepackage{bm}
\usepackage[normalem]{ulem}

\usepackage{color}
\definecolor{Blue}{rgb}{0.00, 0.00, 1.00}
\definecolor{Red}{rgb}{1.00, 0.00, 0.00}
\definecolor{Green}{rgb}{0.00, 0.70, 0.00}

\hypersetup{
    colorlinks=true,       % false: boxed links; true: colored links
    linkcolor=red,          % color of internal links (change box color with linkbordercolor)
    citecolor=blue,        % color of links to bibliography
    filecolor=magenta,      % color of file links
    urlcolor=cyan           % color of external links
}

\newcommand{\be}{\begin{equation}}
\newcommand{\ee}{\end{equation}}
\newcommand{\bea}{\begin{eqnarray}}
\newcommand{\eea}{\end{eqnarray}}

\newcommand{\x}{{\bf x}}

\newcommand{\beq}{\begin{equation}}
\newcommand{\eeq}{\end{equation}}
\newcommand{\beqn}{\begin{eqnarray}}
\newcommand{\eeqn}{\end{eqnarray}}
\begin{document}

\title{
Hole probability for noninteracting fermions in a $d$-dimensional trap}

%\author{Naftali R. Smith}
%\affiliation{LPTMS, CNRS, Univ. Paris-Sud, Universit\'e Paris-Saclay, 91405 Orsay, France}
\author{Gabriel Gouraud}
\affiliation{CNRS-Laboratoire de Physique Th\'eorique de l'Ecole Normale Sup\'erieure, 24 rue Lhomond, 75231 Paris Cedex, France}
\author{Pierre Le Doussal}
\affiliation{CNRS-Laboratoire de Physique Th\'eorique de l'Ecole Normale Sup\'erieure, 24 rue Lhomond, 75231 Paris Cedex, France}
%\author{Satya N. \surname{Majumdar}}
%\affiliation{LPTMS, CNRS, Univ. Paris-Sud, Universit\'e Paris-Saclay, 91405 Orsay, France}
\author{Gr\'egory \surname{Schehr}}
\affiliation{Sorbonne Universit\'e, Laboratoire de Physique Th\'eorique et Hautes Energies, CNRS UMR 7589, 4 Place Jussieu, 75252 Paris Cedex 05, France}
\date{\today}

\begin{abstract}
The hole probability, i.e., the probability that a region is void of particles, is a benchmark of correlations in many body systems.
We compute analytically this probability $P(R)$ for a spherical region of radius $R$ in the case of $N$ noninteracting fermions in their ground state 
in a $d$-dimensional trapping potential. Using a connection to the Laguerre-Wishart ensembles of random matrices, we show that, 
for large $N$ and in the bulk of the Fermi gas, $P(R)$ is described by a universal scaling function of $k_F R$, for which we obtain an exact formula ($k_F$ being the local Fermi wave-vector). It exhibits a super exponential tail $P(R)\propto e^{- \kappa_d (k_F R)^{d+1}}$ where $\kappa_d$ is a universal amplitude, in good agreement with existing numerical simulations. When $R$ is of the order 
of the radius of the Fermi gas, the hole probability is described by a large deviation form which is not universal and which we
compute exactly for the harmonic potential. Similar results also hold in momentum space. 
\end{abstract}

%\pacs{05.40.-a, 02.10.Yn, 02.50.-r}

%05.40.-a: Fluctuation phenomena, random processes, noise, and Brownian motion 
%02.10.Yn	Matrix theory
%02.50.-r	Probability theory, stochastic processes, and statistics 

\maketitle

Since the seminal works of Wigner and Dyson \cite{Dyson,MehtaBook}, the study of the correlations of the eigenvalues of random matrices 
has played a major role in characterizing the statistics of random collections of points, called ``point processes'', beyond
the well known (uncorrelated) Poisson statistics \cite{Brody1981}. Besides the
original applications to the energy level statistics of heavy nuclei \cite{MehtaBook}, more recent applications include 
low dimensional chaotic systems \cite{Haake1991,Chan2018}, mesoscopic disordered conductors \cite{Beenakker1997},
localization/delocalization transitions in disordered quantum systems \cite{Shapiro1993,JS1997,LightSpacing2018,Gritsev2019}. The eigenvalues of
random matrices in the Wigner-Dyson class exhibit level repulsion and spectral rigidity.
A benchmark to quantify this effect is the distribution of the spacing
$s$ between two consecutive eigenvalues, approximately given by the famous Wigner surmise,
i.e. $p(s) \sim s^2 e^{- c s^2}$ for Hermitian random matrices from the Gaussian unitary ensemble (GUE) \cite{MehtaBook,Forrester}.
This spacing distribution can be computed from the {\it hole probability}, i.e. the probability 
that a given interval contains no eigenvalue and it has been calculated for various random
matrix ensembles \cite{BTW,For1992,Mehta1992,Grimm2004,MNSV2011,MMSV14,MMSV16,Forrester}. 

Another interesting example of a random point process is the set of positions of noninteracting fermions
in their ground state in a confining potential. These can be measured in quantum microscopes in 
cold atoms experiments with traps of tuneable shapes \cite{BDZ08,Fermicro1,Fermicro2,Fermicro3,flattrap,Pauli}.
In this case, the randomness originates from quantum fluctuations and the Pauli principle,
which leads to a Slater determinant form for the many body wave-function. It turns out that in one dimension, $d=1$,
this point process can be mapped, in some cases, onto the statistics of the eigenvalues of random matrices.
Indeed, both are examples of determinantal point processes (DPP) on the line
\cite{Eisler1,DeanReview2019}. DPP's are defined by the property that their many body correlations are given by determinants build
from a central object called the kernel \cite{Macchi,Joh_det,Boro_det}. In $d=1$ the hole probability
for fermions can thus often be obtained from random matrix theory (RMT). 

Recently the connections to DPP 
have been much exploited to describe Fermi gases in trapping potentials
in any dimension \cite{DeanEPL2015,DeanPLDReview,SLMS75}. In higher dimension there are some results for the hole probability in two specific examples in $d=2$:
(i) in mathematics, for the zeroes of random series in the complex plane \cite{Sodin2005,Krishnapur2009,Nishry2010}
(ii) in the Ginibre ensemble of random matrices, where there are exact formulae
for the probability that there is no eigenvalue inside a disk \cite{ATW2014,Adhikari2018,lacroix_ginibre}. The latter result can be transposed
in terms of the hole probability for noninteracting fermions in a harmonic trap rotating 
at a critical frequency such that the problem can be mapped to the lowest Landau levels
of a quantum Hall system \cite{LMG19,kulkarni}. 

Besides these two examples, obtaining the hole probability for $d>1$ 
for a general model of trapped noninteracting fermions remains an outstanding question.
In \cite{Torquato2008,Scardicchio2009} this observable was studied numerically and using dimensional arguments 
in the case of free fermions, i.e., in the absence of a trapping potential. 
However there is presently no analytical calculation of this quantity.

In this Letter, we consider noninteracting fermions in $d$ dimensions described
by the single particle Hamiltonian
\be \label{H}
H = \frac{{\bf p}^2}{2 m} + V(r) 
\ee
where $V(r)$ is a central external potential, $r=|{\bf x}|$ and here we work in units such that $m = \hbar = 1$. We 
first obtain, in the case of free fermions (i.e., $V(r)=0$),
an exact formula for the probability $P(R)={\sf P}_d(k_F R)$ that a spherical
domain of radius $R$ contains no fermion (in their ground state). 
Here $k_F=\sqrt{2 \mu}$ is the Fermi wavector and $\mu$ is the Fermi energy.
The scaling function ${\sf P}_d(z)$ can be expressed as a product of Fredholm determinants 
associated to the so-called hard edge Bessel kernel, well known in random matrix theory (RMT),
see Eqs. \eqref{FredBe}, \eqref{eq:forrester}, \eqref{BesselProduct} below. Its asymptotic behavior at small distance is given as $z \to 0$
by
\be \label{small_z_intro}
{\sf P}_d(z) =  1 - B_d z^d + \frac{d}{(d+2)^2} B_d^2 z^{2 d+2}  + O(z^{3d+2},z^{2 d+4}) 
\ee
with $B_d=\frac{1}{2^d \Gamma(1+ \frac{d}{2})^2}$ (see \cite{SM} for higher orders),
which generalizes the result for $d=1$ 
\cite{BTW,Mehta1992,Grimm2004,DeanPLDReview} in which case the level spacing distribution is given by
$p(s) \propto {\sf P}_1''(s)$.

\begin{figure}[t]
\includegraphics[width=\linewidth]{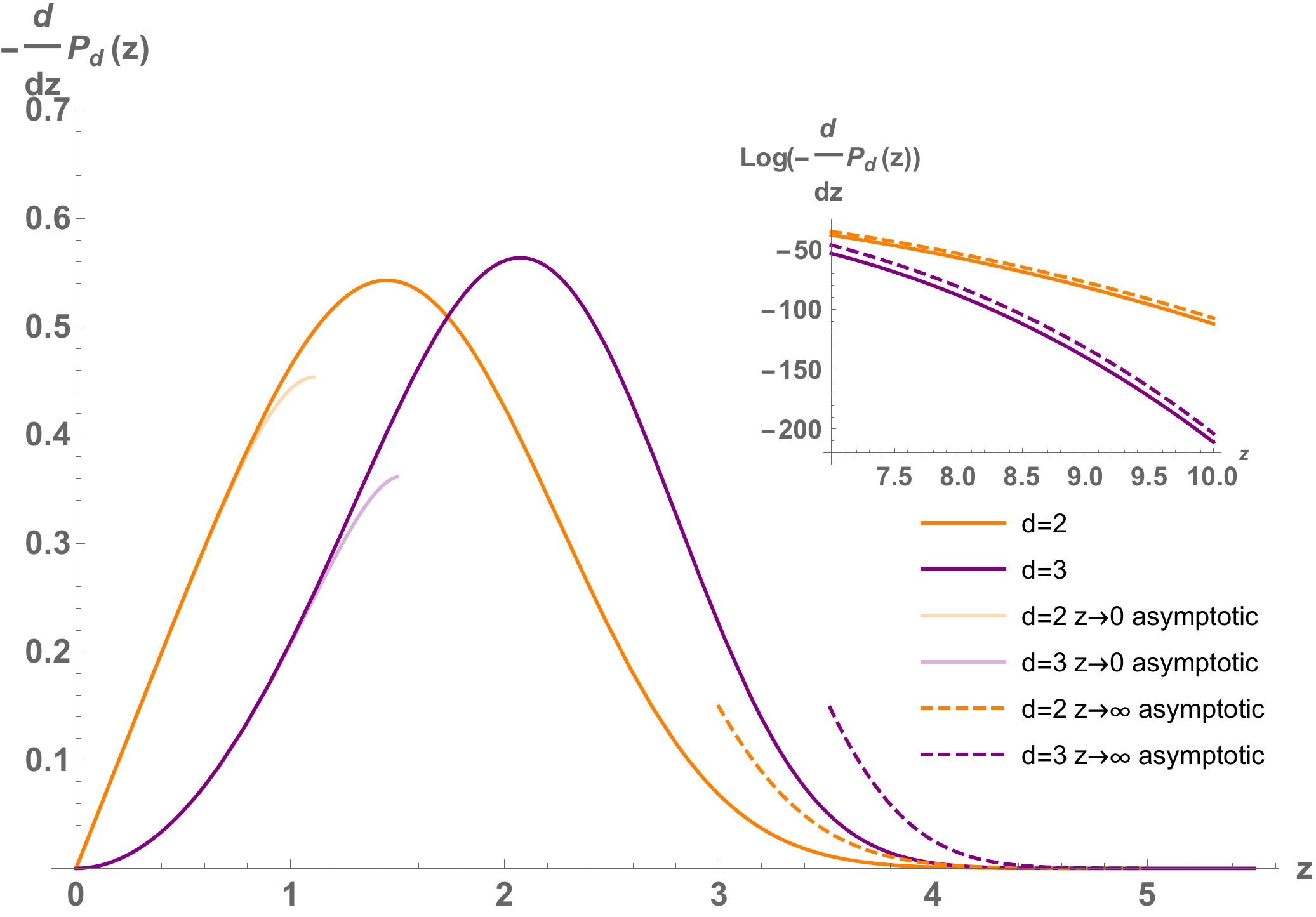}
\caption{Plot of $-\frac{d}{dz}{\sf P}_d(z)$ vs. $z$ in dimension $d=2,3$ from a numerical evaluation (thick line) of \eqref{BesselProduct} using
\eqref{FredBe}. At small $z$ it is well described by the asymptotics (transparent line) from (\ref{small_z_intro}). {\bf  Inset}: same plot
in semi-logarithmic scale, which fits well with the large $z$ super-exponential behavior in (\ref{largez}).}
\label{fig:freefermions}
\end{figure}
At large distance one finds that the hole probability decays super-exponentially
\be \label{largez} 
{\sf P}_d(z) \underset{z \to \infty}{\sim}
\exp( - \kappa_d z^{d+1} ) ~,~ \kappa_d=\frac{2}{(d+1)^2 \Gamma(d+1)} \;.
\ee
This result agrees for $d\to 1$ with $\kappa_1=\frac{1}{2}$ obtained
in \cite{BTW}. The plot of this function ${\sf P}_d(z)$ is
shown in Fig.~\ref{fig:freefermions}.
Note that for $d=2$ the power of the exponential is cubic $z^3$ here, which is at variance
with both the fermion models related to the Ginibre ensemble, and the random
series, for both of which it is $z^4$ 
\cite{ATW2014,Adhikari2018,lacroix_ginibre,Sodin2005,Nishry2010}. We can also compare these results with
the numerical data analysis of \cite{Torquato2008,Scardicchio2009} for free fermions.
In that work the power law $z^{d+1}$ in the exponential was conjectured 
to hold in all dimensions, and the coefficient $\kappa_d$ was measured
numerically. The comparison with our analytic prediction is
presented in Table \ref{comparison}. Although the agreement is quite good
the exact values lie somewhat outside of the error bars, which suggests
that obtaining numerically the true asymptotics requires larger values of $z$.

Next we consider the hole probability for $N$ noninteracting fermions in their ground state and 
in the presence of a smooth confining central potential $V(r)$. We consider here the
limit of large $N$, which corresponds to large Fermi energy $\mu$. In the typical case, for instance 
for a harmonic potential $V(r)=\frac{1}{2} r^2$, 
the fermion density, $\rho(r)$, has a bounded support $r< R_e$ and vanishes beyond the edge at $r=R_e$. 
For large $N$ it is given by the LDA (or semi-classical) expression $\rho(r) = c_d k_F(r)^d$ 
where $k_F(r) = \sqrt{2 (\mu - V(r))_+}$ is the local Fermi wave-vector and $c_d$ is
a constant given below. We consider a spherical region of radius $R$ around the origin.
We find that there are two regimes for the hole probability 
$P(R)$ depending on whether $R$ is of microscopic sizes $R = O(1/k_F(0))$ (typical
interparticle distance), or $R$ is of macroscopic sizes, of order $O(R_e)$. 

In the first regime (microscopic scales) we find that the hole probability takes
the scaling form $P(R) \simeq {\sf P}_d(k_F(0) R)$, where ${\sf P}_d(z)$ is the
same universal scaling function as obtained above for free fermions, with
a non universal scale $1/k_F(0)$. This universality extends to any microscopic 
spherical region located anywhere inside the bulk in the presence of a general
smooth potential. 

In the case of a hole of macroscopic size, $R = O(R_e)$ the probability
$P(R)$ is very small and is characterized by a large deviation form
\be \label{ldev}
P(R) \sim \exp \left(- (k_F(0) R_e)^{d+1} \Psi(\tilde R = R/R_e) \right)
\ee
where the rate function $\Psi(\tilde R)$ is not universal and depends on some
details of the potential $V(r)$. Here we calculate it explicitly in
the case of the harmonic potential $V(r)=\frac{1}{2} r^2$, 
in which case $R_e=\sqrt{2\mu}=k_F(0)$. The function $\Psi(\tilde R)$ 
is given in \eqref{result} and is related to the large deviations in the Wishart-Laguerre ensemble
of random matrices \cite{KC2010}. Its behavior at small argument is found to be 
$\Psi(\tilde R) \simeq \kappa_d \tilde R^{d+1}$ which matches smoothly with
the large distance behavior from microscopic scales, see Eq. \eqref{largez}. 
For large $\tilde R$ it behaves as $\Psi(\tilde R) \sim \tilde R^2$ and
in $d=1$ it is exactly $\Psi(\tilde R) = \frac{1}{2} \tilde R^2$, as found for GUE matrices \cite{MNSV2011,SM}. 

\begin{table}
\begin{tabular}{|c|c|c|}
\hline 
dimension $d$ & numerics \cite{Scardicchio2009} & exact result  \\
\hline
$d=1$ & $0.5$  & $\frac{1}{2}$ \\
%\quad & \quad & \quad \\
\hline
$d=2$ &  $0.1175 \pm 0.0007$    & $\frac{1}{9}=0.1111$  \\
%\quad & \quad & \quad \\
\hline
$d=3$ &  $0.02287 \pm 0.0003$    & $\frac{1}{48} = 0.02083$  \\
%\quad & \quad & \quad \\
\hline
$d=4$ &  $0.00392 \pm 0.00015$    & $\frac{1}{300} = 0.00333$  \\
%\quad & \quad & \quad \\
\hline
\end{tabular}
\caption{Comparison between our exact result \eqref{largez} for $\kappa_d$ (last column) and the numerical estimates of Ref. \cite{Scardicchio2009}}\label{comparison}
\end{table}
%\begin{table}
%\begin{tabular}{|c|c|c|}
%\hline 
%dimension $d$ & numerics \cite{Scardicchio2009} & exact result  \\
%\hline
%$d=1$ & $0.5$  & $\frac{1}{2}$ \\
%\quad & \quad & \quad \\
%\hline
%$d=2$ &  $0.11749 \pm 0.00070$    & $\frac{1}{9}=0.11111$  \\
%\quad & \quad & \quad \\
%\hline
%$d=3$ &  $0.022866 \pm 0.00034$    & $\frac{1}{48} = 0.020833$  \\
%\quad & \quad & \quad \\
%\hline
%$d=4$ &  $0.003923 \pm 0.000145$    & $\frac{1}{300} = 0.003333$  \\
%\quad & \quad & \quad \\
%\hline
%\end{tabular}
%\caption{Comparison between our exact result for $\kappa_d$ (last column) and the numerical estimates of Ref. \cite{Scardicchio2009}}\label{comparison}
%\end{table}
Let us consider $N$ spinless noninteracting fermions in a central potential $V(r)$ in
space dimension $d$. The ground state is obtained as a Slater determinant where all
the eigenstates of the single particle Hamiltonian $\hat H$ in \eqref{H} are occupied up to
the Fermi energy $\mu$. We use the spherical coordinates ${\bf x} = (r,{\bm \theta})$ 
where ${\bm \theta}$ is a $d-1$ dimensional angular vector. 
The Hamiltonian $\hat H$ can be written as 
$\hat{H}=-\frac{1}{2}r^{1-d}\partial_{r}\left(r^{d-1}\partial_{r}\right)+\frac{1}{2r^{2}}\hat{\bm{L}}^{2}+V(r)$, 
and commutes with the angular momentum $\hat {\bm L}$. The eigenfunctions of $\hat H$ thus take the form
$\psi_{n,{\bf L}}(r,{\bm \theta})=r^{\frac{1-d}{2}}\chi_{n,l}(r)Y_{{\bf L}}(\bm{\theta})$
where the $d$-dimensional spherical harmonics $Y_{{\bf L}}(\bm{\theta})$, labeled by the set of angular quantum numbers ${\bf L}$,
are eigenfunctions of $\hat{\bm{L}}^2$ with eigenvalues $\ell(\ell+d-2)$, $\ell=0,1,\dots$,
which defines the angular sector.
The radial parts $\chi_{n,\ell}(r)$ are the eigenfunctions of a collection of 1D radial Hamiltonians
$\hat H_\ell = - \frac{1}{2} \partial_r^2  + V_\ell(r)$, $r \geq 0$, with potentials
%\cite{MoshinskyBook,Farthest} 
\be \label{Vl} 
V_\ell(r)= V(r) + \frac{a^2- \frac{1}{4} }{2 r^2} \quad , \quad a=\ell+ \frac{d}{2}-1
%V_\ell(r)= V(r) + \frac{(\ell + \frac{d-3}{2})(\ell + \frac{d-1}{2})}{2 r^2} 
\ee 
with eigenenergies $\epsilon_{n,\ell}$, $n=0,1,\dots$, each with degeneracy
$g_d(\ell) = \frac{(2 \ell + d-2) \Gamma(\ell+d-2)}{\Gamma(\ell+1) \Gamma(d-1)}$, $\ell \geq 1$
and $g_d(0)= 1$. In the ground state of the $N$ fermions, each angular sector 
the lowest $m_\ell$ energy levels are occupied, i.e $n=0,\dots, m_\ell-1$, such
that $\epsilon_{n,\ell} \leq \mu$ and $N = \sum_{\ell} g_d(\ell) m_\ell$. 

Let us now focus on the example of the harmonic oscillator, $V(r)= \frac{1}{2} r^2$.
In that case the eigenfunctions of \eqref{Vl} can be computed exactly, and are given by Laguerre polynomials
$\chi_{n,\ell}(r) \propto r^{a+ \frac{1}{2}} {\rm L}_n^{a}(r^2) e^{-r^2/2}$
with eigenenergies $\epsilon_{n,\ell}=2 n + a + 1$, where $a= \ell+ \frac{d}{2}-1$.
The number of occupied states in the ground state within the $\ell$ sector is thus $m_\ell = {\rm Int}(\frac{\mu-\ell-{d}/{2}}{2} + 1)$, 
where ${\rm Int}(z)$ denotes the integer part of $z$. Note that
$m_\ell=0$ for $\ell > \ell_{\max}(\mu)= \mu- \frac{d}{2}$ (where $\mu$ is integer
for even $d$ and half-integer for odd $d$). The ground state wavefunction is given by  
$\Psi_0({\bf x_1}, \cdots, {\bf x_N}) = \frac{1}{\sqrt{N!}} \det_{1\leq i,j \leq N} \left[ \psi_{{\bf k}_i}({\bf x}_j) \right]$,
where ${\bf k}_i = (n_i, {\bf L}_i)$ labels the single particle eigenfunction of the occupied eigenstates. We assume here
that the ground state is non-degenerate (i.e., the last level is fully occupied,
see discussion in \cite{Farthest}). 

We now compute the hole probability $P(R)$ as the probability that there is no fermion in the
sphere of radius $R$ centered on the origin. It is given by
\be \label{holedef} 
P(R)= \prod_{i=1}^N  \int_{|{\bf x}_i|>R} d^d {\bf x}_i \; |\Psi_0({\bf x_1}, \cdots, {\bf x_N})|^2 \;.
\ee
Using the Cauchy-Binet formula (see e.g. \cite{Farthest,CalabreseMinchev1}) it can be written
as a determinant 
\bea \label{fd1} 
P(R) = \det_{1 \leq i,j \leq N} \left[\delta_{ij} - \mathbb A_{ij} \right]
\eea 
in terms of the overlap matrix $\mathbb A_{ij} = \int_{r = |\x| \leq R} d^d{\bf x} \, \psi^*_{n_i, {\bf L}_i}({\bf x}) \psi_{n_j,{\bf L}_j}({\bf x})$.
Using the orthogonality of the spherical harmonics, the angular integration gives 
$\mathbb A_{ij} = \delta_{{\bf L}_i, {\bf L}_j} \mathbb A^{(\ell)}_{ij}$ 
with $\mathbb A^{(\ell)}_{ij}=\int_{0}^{R} dr \, \chi_{n_i,\ell_i}(r) \chi_{n_j,\ell_i}(r)$.
%where we recall that $\int_0^\infty  dr\ \chi_{kl}(r)\chi_{k'l}(r)=\delta_{kk'}$. 
Hence the matrix $\mathbb{A}$ is diagonal in the variables ${\bf L}_i$, and the determinant factorises
over the angular sectors \cite{Farthest}
\be
P(R) = \prod_{\ell=0}^{\ell_{\rm max}(\mu)} P_\ell(R)^{g_d(\ell)} ~,~
\quad P_\ell(R)= \det_{1 \leq i,j \leq m_\ell} \left[\delta_{ij} - \mathbb A^{(\ell)}_{ij} 
\right] \label{product} 
\ee 
where $P_\ell(R)$ is the probability that the interval $[0,R]$ is empty in the ground state
of $m_\ell$ noninteracting fermions described by the single particle Hamiltonian $\hat H_\ell$.
Note that formula \eqref{product}, \eqref{fd1}, \eqref{holedef} are valid for any central potential $V(r)$.

We will now start by studying the 1D radial problem to obtain $P_\ell(r)$ within each
$\ell$ sector, and in a second stage we will evaluate the product \eqref{product}.
For a general potential $V(r)$ this is a difficult problem, 
however in the case of the harmonic oscillator we can make further progress by using a connection to
the complex Wishart-Laguerre (WL) ensemble of random matrix theory \cite{Forrester,Dumitriu2002,NadalMajumdar2009}. It is defined by the
following joint probability distribution function (PDF) for a set of $m$ eigenvalues $\lambda_i$
\be  \label{WL} 
P^{(m)}_{\rm WL}(\vec \lambda) \propto e^{- \sum_{i=1}^m \lambda_i} \prod_{i=1}^m \lambda_i^{\nu} 
\prod_{1 \leq j,k \leq m} (\lambda_j -\lambda_k)^2
\ee
which depends on the continuous parameter $\nu >-1$. In the case where $\nu$ is a positive integer,
it describes the eigenvalues of complex Wishart 
random matrices. They are of the form $W=X^T X$ where $X$ is a $M \times m$ rectangular random
matrix with i.i.d. unit complex Gaussian entries with $M \geq m$
(see \cite{VivoMajumdar} for the case $M<N$), and in that case $\nu=M-m$.
%The joint probability distribution function (PDF) of the eigenvalues 
%of $W$ reads 
%\be  \label{WL} 
%P^{(m)}_{\rm WL}(\vec \lambda) \propto e^{- \sum_{i=1}^m \lambda_i} \prod_{i=1}^m \lambda_i^{M-m} 
%\prod_{1 \leq j,k \leq m} (\lambda_j -\lambda_k)^2
%\ee
Consider now the ground state wave function of $m_\ell$ noninteracting fermions in one dimension, in the potential 
$V_\ell(r) = \frac{1}{2} r^2 + \frac{a^2- \frac{1}{4}}{2 r^2}$, associated to the radial problem \eqref{Vl}
with $a=\ell+\frac{d}{2}-1$. 
It is given by the Slater determinant $\Phi_0(\vec r) = \frac{1}{\sqrt{m_\ell!}} \det_{1\leq i<j \leq m_\ell} \left[ \chi_{i-1,\ell}(r_j) \right]$,
with $\vec r=(r_1,\dots,r_{m_\ell})$. Remarkably, the quantum joint PDF of the fermion positions $r_i$
is identical to the joint PDF of the WL eigenvalues $\lambda_i$ \eqref{WL} upon the correspondence $\lambda_i=r_i^2$,
$m=m_\ell$ and $a=\ell + \frac{d}{2}-1 =\nu$, i.e., one has $|\Phi_0(\vec r)|^2 \prod_{i=1}^{m_\ell} dr_i =  P^{(m_\ell)}_{\rm WL}(\vec \lambda) \prod_{i=1}^{m_\ell} d\lambda_i$
(see \cite{Farthest} for details).

The calculation of $P_\ell(R)$ is then equivalent to computing the cumulative distribution function (CDF) of
the smallest eigenvalue $\lambda_{\min}$ in the WL ensemble, i.e.,  $P_\ell(R)= {\rm Prob}(\lambda_{\min} >R^2)$.
This CDF was studied in several works \cite{TWB1994,FH1994,SM2010,Ehr2010,EdelmanGuionnetPeche2016,PS2016} in RMT. 
In particular in \cite{FH1994} a useful determinantal formula was found in the case where $\nu$ is a positive
integer. Translated into the fermion problem in the case of $d$ an even integer, $P_\ell(R)$ is 
given by a $\nu \times \nu$ determinant (with $\nu=a=\ell+ \frac{d}{2}-1$). Together with \eqref{product}
it gives an exact formula for the hole probability for $N$ fermions in the $d$-dimensional harmonic potential,
displayed in Eqs. \eqref{product2} and \eqref{HOhole} in \cite{SM}. It allows to
plot the exact hole probability for a small number of fermions. 

We now consider the limit of a large number of fermions, $N \gg 1$, equivalently large $\mu$. In this limit
there are two distinct regimes, microscopic and macroscopic, as mentioned in the introduction. The mean density of fermions is given
by the LDA formula $\rho(r)= c_d k_F(r)^d $, with $c_d=\frac{1}{2^d \pi^{d/2} \Gamma(1+d/2)}$ and $k_F(r)=\sqrt{2 \mu - r^2}$
for the harmonic oscillator \cite{DeanPLDReview}. Near the center of the trap, the typical distance between particles is thus $1/k_F(0)=1/\sqrt{2 \mu}$, which defines
the microscopic scale. On the other hand, the edge of the Fermi gas is a sphere of macroscopic radius $R_e=\sqrt{2\mu} \gg 1/k_F(0)$.

{\it Microscopic regime.} This corresponds to the case where $R = O(\frac{1}{\sqrt{\mu}})$ is of order
of the interparticle spacing at $r=0$. Coming back to the decomposition in angular sectors, 
since $\mu$ is large in each sector the number of fermions $m_\ell \simeq (\mu-\ell)/2$ is also large.
It is easy to see that only sectors with $\ell = O(1)$ contribute to the product in \eqref{product}.
Indeed, the centrifugal energy $(a^2-1/4)/(2 r^2)$ in $V_\ell(r)$ in \eqref{Vl} must remain 
at most of $O(\mu)$, and therefore for $r \sim 1/\sqrt{\mu}$ only values of 
$a = O(1)$ are allowed. Hence one can approximate $m_\ell \simeq \mu/2$
for all $\ell$ in that regime. Note that the potential term $V(r)=\frac{1}{2} r^2$
is negligible in that regime, hence it is identical to free fermions. 

This regime corresponds in the RMT context to the so-called hard edge scaling regime.
In that regime the $m_\ell$ eigenvalues of the WL ensemble of parameter $\nu=a$ \eqref{WL},
with $m_\ell$ large,
are of order $\lambda_i = O(1/m_\ell)$. More precisely, the scaled eigenvalues $b_i = 4 m_\ell \lambda_i$ form a DPP,
called the Bessel process of order $\nu=a=\ell + d/2-1$, described by the kernel
\be \label{kernelbessel}
%\lambda_i = \frac{b_i}{4 m_{\ell}}
K^B_\nu(b,b') = \frac{1}{4}  \int_0^1 dz \, J_{\nu}(\sqrt{z b}) J_{\nu}(\sqrt{z b'}) \;,
%= \frac{\sqrt{b'} J'_\ell(\sqrt{b'}) J_\ell(\sqrt{b}) - \sqrt{b} J'_\ell(\sqrt{b}) J_\ell(\sqrt{b'}) }{2 (b-b') } 
\ee 
where $J_\nu(z)$ is the Bessel function of index $\nu$. Using standard results for DPP the hole probability in the $\ell$ sector is given as a Fredholm
determinant \cite{Joh_det,Boro_det}
\be \label{FredBe}
F_\nu(b):={\rm Prob}( b_{\min} > b) = {\rm Det}(I - P_{[0,b]} K^B_\nu) \;,
\ee
where $b_{\min} = \min_i b_i$ and $P_{[0,b]}$ is the projector on the interval $[0,b]$. This FD can be expressed as
$\log F_\nu(b)= - \int_0^b \frac{ds\, \sigma(s)}{s}$ from the solution
$\sigma(s)$ of the Painlev\'e III equation \cite{TWB1994}
\be
(s \sigma'')^2 + \sigma' (\sigma - s \sigma')(4 \sigma'-1) - \nu^2 (\sigma')^2 = 0   \;,
\ee
where $\sigma(s) \simeq \frac{s^{1+\nu}}{2^{2 \nu+2} \Gamma(1+\nu) \Gamma(2+\nu)}$
at small $s$. For even space dimension $d$, i.e., integer $\nu$, as discussed above, there are other representations for the hole probability,
which in this microscopic regime lead to the remarkably simple formula \cite{FH1994} 
\be \label{eq:forrester}
F_\nu(b) = e^{- b/4} \det_{1 \leq j,k \leq \nu} I_{j-k}(\sqrt{b}) \;, %= e^{- b/4}  \int_{U(\ell)} e^{\frac{1}{2} \sqrt{b} {\rm Tr} (U + U^+)} 
\ee
where $I_n(x)$ is the modified Bessel function. Using formula \eqref{product}, we obtain that 
in this scaling regime, the hole 
probability $P(R)$ for the fermions in $d$ dimensions, takes the 
scaling form 
$P(R) \simeq {\sf P}_d(k_F(0) R)$, where the scaling function is given as an infinite product
\be \label{BesselProduct} 
{\sf P}_d(z) =  \prod_{\ell=0}^{+\infty}  F_{\ell+ \frac{d}{2}-1} (z^2)^{g_d(\ell)} \;.
\ee 
This result, which we derived for the harmonic oscillator (with $k_F(0)=\sqrt{2 \mu}$) holds asymptotically for large $N$ for any smooth trapping potential.
In addition, it is {\it exact} for free fermions in $d$ dimensions, with $k_F=\sqrt{2 \mu}$. 
Note that for free fermions, an alternative formula exists using the $d$-dimensional
extension of the sine-kernel \cite{DeanPLDReview,Torquato2008}, 
$K_d({\bf r}, {\bf r}') = \frac{J_{d/2}(|{\bf r}- {\bf r}'|) }{(2 \pi |{\bf r}- {\bf r}'|)^{d/2} }$.
It is given by another Fredholm determinant 
\be \label{Besselddim} 
{\sf P}_d(z) =  {\rm Det}(I - P_z K_d) 
\ee 
where $P_z$ is the projector on $|{\bf r}|<z$. The formulae \eqref{BesselProduct} and \eqref{Besselddim} are
in fact {\it equivalent} (which is not a trivial property). Both formulae can be expanded in small $z$,
leading to \eqref{small_z_intro} and pushed to higher orders in \cite{SM}.
While the expansion of \eqref{Besselddim} is straightforward, expanding \eqref{BesselProduct} requires
to solve the Painlev{\'e} III equation at small arguments. The formula \eqref{BesselProduct} however
allows to study the asymptotic behavior of ${\sf P}(z)$ at large $z$, as we now show.

In the large $z$ limit one needs the asymptotics of $F_\nu(b)$ at large $b$. 
One can check \cite{SM} that the asymptotics of the infinite product in 
\eqref{BesselProduct} is dominated by large values of $\ell$, for which 
the decay of $F_\nu(b)$ occurs on scale $b \sim \ell^2$. This double limit for \eqref{eq:forrester}
was studied, using Coulomb gas techniques, in the context of lattice QCD in \cite{GW1980,Wadia} and later in the study of the longest increasing subsequence of random permutations \cite{Joh1998} (see also \cite{KC2010})
and it was shown to take the scaling form (with $\nu \sim \ell$)
\bea \label{phi_p}
&& F_\nu(b) \sim \exp\left[ - \ell^2 \phi_+\left(\gamma= \frac{\sqrt{b}}{\ell} \right) \right] \;, \\
&& \phi_+(\gamma) = \theta(\gamma-1) ( \frac{\gamma^2}{4} -\gamma + \frac{1}{2} \log \gamma + \frac{3}{4} ) \;. \nonumber
\eea 
Inserting this expression into \eqref{BesselProduct}, approximating the sum over $\ell$ by an integral, 
using that $g_d(\ell) \simeq \frac{2 \ell^{d-2}}{\Gamma(d-1)}$ at large $\ell$, one obtains 
\bea\label{Pdl}
{\sf P}_d(z) \sim \exp\left[ - \frac{2}{\Gamma(d-1)} \int_0^{+\infty} d\ell \ell^d \phi_+\left(\frac{z}{\ell}\right) \right]
\eea 
leading to our main result \eqref{largez}, with 
$\kappa_d = \frac{2}{\Gamma(d-1)}
 \int_1^{+\infty} \frac{d\gamma}{\gamma^{d+2}}  \phi_+(\gamma) = \frac{2}{(d+1)^2 \Gamma (d+1)}$.
 Note that the calculations of \cite{GW1980,Joh1998} use the formula 
 \ref{eq:forrester} valid only for even $d$, however one can
 also obtain \eqref{phi_p} in any $d$ using the Painlev{\'e} equation \cite{SM}.

{\it Macroscopic regime.} We now focus on the harmonic potential $V(r)=\frac{1}{2} r^2$, in the large $N$ limit, in
which case the density has an edge at $r=R_e=\sqrt{2 \mu}$. The macroscopic regime corresponds $R = O(R_e) = O(\sqrt{\mu})$.
Since $\mu$ is large, in each angular sector the number of fermions is again $m_\ell \simeq (\mu-\ell)/2$, however in this regime the product in \eqref{product} is controlled by the values of $\ell = O(\mu)$.
In the language of the WL ensemble \eqref{WL} this corresponds to the limit of large matrix size $m=m_\ell \to +\infty$,
and large index $\nu=a=\ell+ \frac{d}{2}-1 \to +\infty$, with 
with $\alpha = \frac{\nu}{m}$ fixed of order unity. Using the correspondence with fermions discussed above, $\lambda_i=r_i^2$,
one has $\alpha = \frac{\nu}{m} \simeq \frac{\ell}{m_\ell}=O(1)$. It is known that in this regime the spectrum of the
WL matrices has support on the interval $[\lambda_-,\lambda_+]$ with $\lambda_{\pm} = m \zeta_\pm$
and $\zeta_\pm = (1 \pm \sqrt{1+\alpha})^2$. The correspondence with fermions shows that
within each $\ell$ sector, the mean fermion density $\rho_\ell(r)$ has support 
$[r_-(\ell), r_+(\ell)]$ with 
\be \label{rell} 
r_\pm(\ell)^2 = \lambda_{\pm} = m_\ell \zeta_{\pm} \simeq \mu \pm \sqrt{\mu^2 - \ell^2}
\ee
which coincides with the result obtained using the LDA, i.e.,   
$\rho_\ell(r)=\frac{1}{\pi} \sqrt{2 (\mu - V_\ell(r))_+}$ where $V_\ell(r)$
is given in \eqref{Vl}. 

For each sector $\ell$, there are a priori three different scaling regimes for $P_\ell(R)={\rm Prob}(\lambda_{\rm \min}> R^2)$
when $R=O(\sqrt{\mu})$ and $\ell=O(\mu)$. Indeed, it is known that the smallest eigenvalue
$\lambda_{\rm min}$ of a WL random matrix (\ref{WL}) exhibits three regimes \cite{SM2014}:

(i) A typical fluctuation regime around the lower
edge for $\lambda_{\rm min} - \lambda_- = O(m^{1/3})$, described by the "soft-edge" Tracy-Widom distribution $F_2$.
This regime will not play a role here.

(ii) A "pulled" large deviation regime to the left of $\lambda_-$, i.e., $\lambda_{\rm min} - \lambda_- = O(m) <0 $,
which in terms of the fermions read \cite{SM2014}
$P_{\ell}(R)   \sim 1- e^{- 2 m_\ell \Phi_-\left(\frac{m_\ell \zeta_--R^2}{m_\ell}, \frac{\ell}{m_\ell}\right) }$.

(iii) A "pushed" large deviation regime to the right of $\lambda_-$, i.e., $\lambda_{\rm min} - \lambda_- = O(m^2) >0 $,
which in terms of the fermions read
\be \label{regime3} 
P_{\ell}(R)  \sim e^{- 2 m_\ell^2 \Phi_+\left(\frac{R^2- m_\ell \zeta_-}{m_\ell}, \frac{\ell}{m_\ell}\right) } \;.
\ee
The rate functions $\Phi_\pm(z,\alpha)$ were computed in Ref. \cite{KC2010} using Coulomb gas methods
and are recalled in \cite{SM}. 

Consider now the expression for the logarithm of the hole probability $\log P(R)$ expressed as a sum over $\ell$ from \eqref{product}. 
As $\ell$ increases in $[0,\ell_{\max}(\mu) \simeq \mu]$, the edge $r_-(\ell)$ in \eqref{rell}
%= m_\ell \zeta_{-} \simeq \mu \pm \sqrt{\mu^2 - \ell^2}$
increases from $0$ to $\sqrt{\mu}$. There are two cases. For $R^2>\mu$ one has $R>r_-(\ell)$ for all $\ell<\mu$, and only the regime
(iii) applies. For $R^2<\mu$, a priori the three regimes apply, i.e., regime (ii) for the sectors with $\ell>\ell_+(R)=2 R \sqrt{2 \mu - R^2}$ 
and regime (iii) for the sectors $\ell<\ell_+(R)$. However, one sees that regime (ii) contributes only
to exponentially small corrections. Hence the leading contributions come from regime (iii) in \eqref{regime3} 
in all cases. 
In computing the logarithm of \eqref{product} we can approximate the
sum over $\ell$ by an integral and $g_d(\ell) \simeq \frac{2 \ell^{d-2}}{\Gamma(d-1)}$.
Performing the change of variable $v=\ell/\mu$ we obtain the large deviation formula for the hole probability in the form \eqref{ldev} 
with $k_F(0) R_e= 2 \mu$ and $\tilde R=R/\sqrt{2\mu}$
with the rate function
\bea \label{result} 
&& \Psi(\tilde R) = 
\int_0^{v_{\rm max}(\tilde R)} \frac{v^{d-2} (1 - v)^2 dv}{2^{d+1} \Gamma(d-1)} \\
&& \times \,
  \Phi_+\left(\frac{4 \tilde R^2}{1-v} -  (1 - \sqrt{1 + \alpha_v})^2, 
\alpha_v\right) \nonumber 
\eea 
where $\alpha_v = \frac{2 v}{1-v}$ and 
$v_{\rm max}(\tilde R) = 2 \tilde R \sqrt{1- \tilde R^2}$ for $\tilde R^2<1/2$
and $v_{\rm max}(\tilde R) = 1$ for $\tilde R^2>1/2$. 
The function $\Phi_+(z,\alpha)$ being quite complicated, the integral in \eqref{result} has been evaluated
numerically in \cite{SM}. It exhibits a transition of high order at $\tilde R=1/\sqrt{2}$
\cite{footnotetransition} and its asymptotic behaviors can be extracted \cite{SM}. 
%In particular, 
%for small $\tilde R \ll 1$ one finds 
%and change the integration variable from $v$ to $\alpha=\alpha_v$
%leads to 
%\be
%\Psi(\tilde R) \simeq \int_0^{4 \tilde R} \frac{\alpha^{d-2}  d\alpha}{2^{2d} \Gamma(d-1)}
%\Phi_+((\alpha^2( 4 (\frac{\tilde R}{\alpha})^2 - \frac{1}{4}) , \alpha)   
%\ee 
%Using that for $\alpha \to 0$, $\Phi_+(\alpha^2 y,\alpha) \simeq \frac{\alpha^2}{2} \phi_+(\gamma= \sqrt{1+ 4 y})$
%\cite{SM}, one finds that for small $\tilde R \ll 1$, $\Psi(\tilde R) 
%\simeq \frac{1}{2^{2 d+1} \Gamma(d-1)} \int_0^{4 \tilde R} d \alpha \alpha^d \phi_+(\gamma = \frac{4 \tilde R}{\alpha}) 
%= \kappa_d \tilde R^{d+1}$, hence we find 
%$\Psi(\tilde R)  \simeq \kappa_d \tilde R^{d+1}$ 
%which matches exactly with the large radius limit of the microscopic regime Eq.~\eqref{largez}. 

%{\red Maybe we should also give (i) the next order at small $\tilde R$ (ii) the large $\tilde R$ behavior}

In conclusion, we have computed analytically the hole probability for noninteracting
fermions in a $d$-dimensional central trapping potential. We have obtained an exact formula for the universal
scaling function ${\sf P}_d(z)$ which describes holes of size of the order of interparticle distance, in an arbitrary
smooth potential. The asymptotics of our results are in good agreement with existing numerical simulations \cite{Torquato2008,Scardicchio2009}.
It characterizes the rigidity of the Fermi gas, a generalization of level repulsion in random matrix theory. 
In addition we have obtained, for the harmonic oscillator the full large deviation function
for macroscopic holes. Interestingly, our results also apply to the hole probability in momentum space \cite{SM} which
can be measured from time of flight experiments \cite{flattrap}. The method introduced here could allow
to predict a larger variety of probes of fermion correlations in traps \cite{Pauli}. It is also possible
to incorporate finite temperature effects, and we hope that our results can be compared
with cold atom experiments \cite{Pauli}. It would be interesting to also study the hole
probability for interacting systems for which very few results exist,
mostly in the related context of spin chains \cite{Kitanine,Morin,KorepinBook}.

%$c_d=\frac{1}{2^d \pi^{d/2} \Gamma(1+d/2)}$

%\begin{figure}[ht]
%\centering
%\includegraphics[angle=0,width=1.0\linewidth]{NumberVariance2dHO_mu100.pdf}
%
%\caption{Variance of ${\cal N}_{\cal D} = {\cal N}_R$ for a disk of radius $R$ in $d=2$, plotted vs $\tilde R = R/\sqrt{2 \mu}$ for $\mu = 100$.  
%The simulations (symbols) show excellent agreement with our predictions: In the bulk, with 
%(\ref{variance_N_ddimHO}) (solid line), which includes (\ref{Ad}) and 
%$B_2(\tilde R)$ in 
%%TODO: make sure that this equation number remains correct in all future revisions
%Eq.~(82) in \cite{SM}, and near the edge $\tilde R=1$, with the scaling form (\ref{final})  (dotted line).
%%
%{\bf Inset:} the sub-leading term $B_2(\tilde R)$ %in Eq. (80) in \cite{SM}, 
%plotted vs $\tilde R$ (yellow dashed line), %in excellent agreement with 
%compared to the simulations (symbols), the leading term $A_2(\tilde R) \mu \log \mu$ being subtracted from the variance.}
%
%\label{Fig_density}
%\end{figure}

{\it Acknowledgments:} 
We thank D. S. Dean, S. N. Majumdar and N. R. Smith for useful discussions on closely related topics. 
This research was supported by ANR grant ANR-17-CE30-0027-01 RaMaTraF.

{}

\newpage

.

\newpage

\begin{widetext} 

%%%%%%%%%%% Merge with supplemental materials %%%%%%%%%%
%\pagebreak
%%\widetext
%\begin{widetext} 
%\begin{center}
%\textbf{\large Supplemental Materials: Title for main text}
%\end{center}
%%%%%%%%%%% Merge with supplemental materials %%%%%%%%%%
%%%%%%%%%%% Prefix a "S" to all equations, figures, tables and reset the counter %%%%%%%%%%
%\setcounter{equation}{0}
%\setcounter{figure}{0}
%\setcounter{table}{0}
%\setcounter{page}{1}
%\makeatletter
%\renewcommand{\theequation}{S\arabic{equation}}
%\renewcommand{\thefigure}{S\arabic{figure}}
%\renewcommand{\bibnumfmt}[1]{[S#1]}
%\renewcommand{\citenumfont}[1]{S#1}
%%%%%%%%%%% Prefix a "S" to all equations, figures, tables and reset the counter %%%%%%%%%%

%\setcounter{section}{0}
%\renewcommand{\thesection}{S-\Roman{section}}

\setcounter{secnumdepth}{2}

\begin{large}
\begin{center}

Supplementary Material for\\  {\it Hole probability for noninteracting fermions in a $d$-dimensional trap}

\end{center}
\end{large}

\bigskip

We give the principal details of the calculations described in the main text of the Letter. We present some related numerical results. 

\bigskip

\section{Comparison with the results of Refs. \cite{Torquato2008,Scardicchio2009} }

In Refs. \cite{Torquato2008,Scardicchio2009} the hole probability for free fermions in $d>1$ dimensions was studied, by
dimensional arguments and numerical simulations. To compare with our results, let us recall
that the authors of Refs. \cite{Torquato2008,Scardicchio2009} write the hole probability as  
\be
P(R) = e^{- \rho \frac{2 \pi^{d/2}}{\Gamma(d/2)} \int_0^R x^{d-1} G_V(x)\, dx } 
\ee 
and numerically determine the large $x$ behavior of the function $G_V(x) \sim \alpha_d \, x$ (fitting with an assumed linear function). 
The coefficient $\alpha_d$ is given the Table I in \cite{Scardicchio2009}. On the other hand, our prediction reads, at large $R$
\be
P(r) = e^{- \kappa_d (k_F R)^{d+1} }
\ee 
The relation between $k_F$ and $\rho$ is
\be
\rho=\frac{k_F^d 2^{-d}}{\pi^{d/2} \Gamma(1+ \frac{d}{2}) } 
\ee
Since they use $\rho=1$ their predicted values for $\kappa_d$ are
\be
\kappa_d^{\rm num} = \frac{\alpha_d 2^{-d} \Gamma \left(\frac{d}{2}+1\right)^{-\frac{d+1}{d}}}{\sqrt{\pi }
   (d+1) \Gamma \left(\frac{d}{2}\right)}
\ee
from which we have constructed the Table \ref{comparison} in the Letter.

\section{Hole probability from the Bessel process}

In the text we have shown that the hole probability for fermions in a central potential $V(r)$ in dimension $d$ can be written
as the product \eqref{product} over factors $P_\ell(R)$ corresponding to angular sector $\ell$. Each factor $P_\ell(R)$ equals to the hole probability 
for $m_\ell$ fermions of positions $r_i \in \mathbb{R}^+$ in the one-dimensional potential $V_\ell(r) = V(r) + \frac{a^2 - \frac{1}{4}}{2 r^2}$ in \eqref{Vl} 
with $a=\ell+ \frac{d}{2}-1$, at Fermi energy $\mu$.
This means that $P_\ell(R) = {\rm Prob}( r_{\rm min} > R)$, where $r_{\min}$ is the position of the leftmost
fermion on the positive half-axis. For simplicity, we now specialize to the harmonic potential $V(r)= \frac{1}{2} r^2$,
although, as mentioned in the text, the results below are valid for more general smooth potentials, i.e., $V(r) \propto r^p$ with $p>0$, upon the change $k_F = \sqrt{2 \mu} \to k_F(0)=\sqrt{2 \mu- V(0)}$. 
In that case one can use the mapping in Eq. \eqref{WL} to the WL ensemble of random matrices of eigenvalues $\lambda_i=r_i^2$ and
one finds that $P_\ell(R) = {\rm Prob}( \lambda_{\rm min} > R^2$), with $\lambda_{\min}= \min_{i} \lambda_i$. 

Consider now the limit of large $N$, i.e., large $\mu$. In the microscopic regime, $R \sim \frac{1}{\sqrt{\mu}}$, the values of $\ell$ which dominate the product \eqref{product}
are $\ell = O(1)$. Therefore we now focus on these values of $\ell$, for which $m_\ell \simeq \mu/2$, hence the number of fermions in
each sector $m_\ell$ is large. Since
$R \sim \frac{1}{\sqrt{\mu}}$, it means that we are looking at $\lambda_{\rm min} = O(\frac{1}{\mu})=O(\frac{1}{m_\ell})$.
This the so-called universal hard-edge regime of the WL ensemble, where it is known that the eigenvalues $\lambda_i$ behave, for large $m_\ell$ as
\be
\lambda_i = r_i^2 \simeq \frac{b_i}{ 4 m_\ell} 
\ee 
where the $b_i = O(1)$ form a determinantal point process (DPP), called the Bessel process of index $\nu=a=\ell + \frac{d}{2}-1$, described by the kernel \eqref{kernelbessel} \cite{Forrester,TW1994}. 
Since the hole probability of the Bessel process can be written as a Fredholm
determinant \cite{Joh_det,Boro_det} one obtains 
\be \label{PFD} 
P_\ell(R) \simeq F_\nu(b)={\rm Prob}( b_{\min} > b) = {\rm Det}(I - P_{[0,b]} K^B_\nu) \quad , \quad b = 4 m_\ell R^2 = (k_F R)^2 = z^2 
\ee
where $k_F=\sqrt{2 \mu}$ and $b_{\min} = \min_i b_i$ is the position of the left-most point of the Bessel process (corresponding to the smallest eigenvalue in the WL ensemble). The asymptotics of $F_\nu(b)$ are $1-F_\nu(b) \sim b^{\nu+1}$ at small $b$, and 
at large $b$ (and fixed $\nu$) one has (see formula 1.24 in \cite{TWB1994})
\be
\log F_\nu(b) \simeq - \frac{b}{4} + \nu \sqrt{b} - \frac{\nu^2}{4} \log b 
+ \log(\frac{G(1+\nu)}{(2 \pi)^{\nu/2}}) + o(1) 
\ee
More detailed small $b$ asymptotics are performed in Section \ref{sec:pain}. 
\\
\begin{figure}[t]
    \centering
    \includegraphics[width=0.8\linewidth]{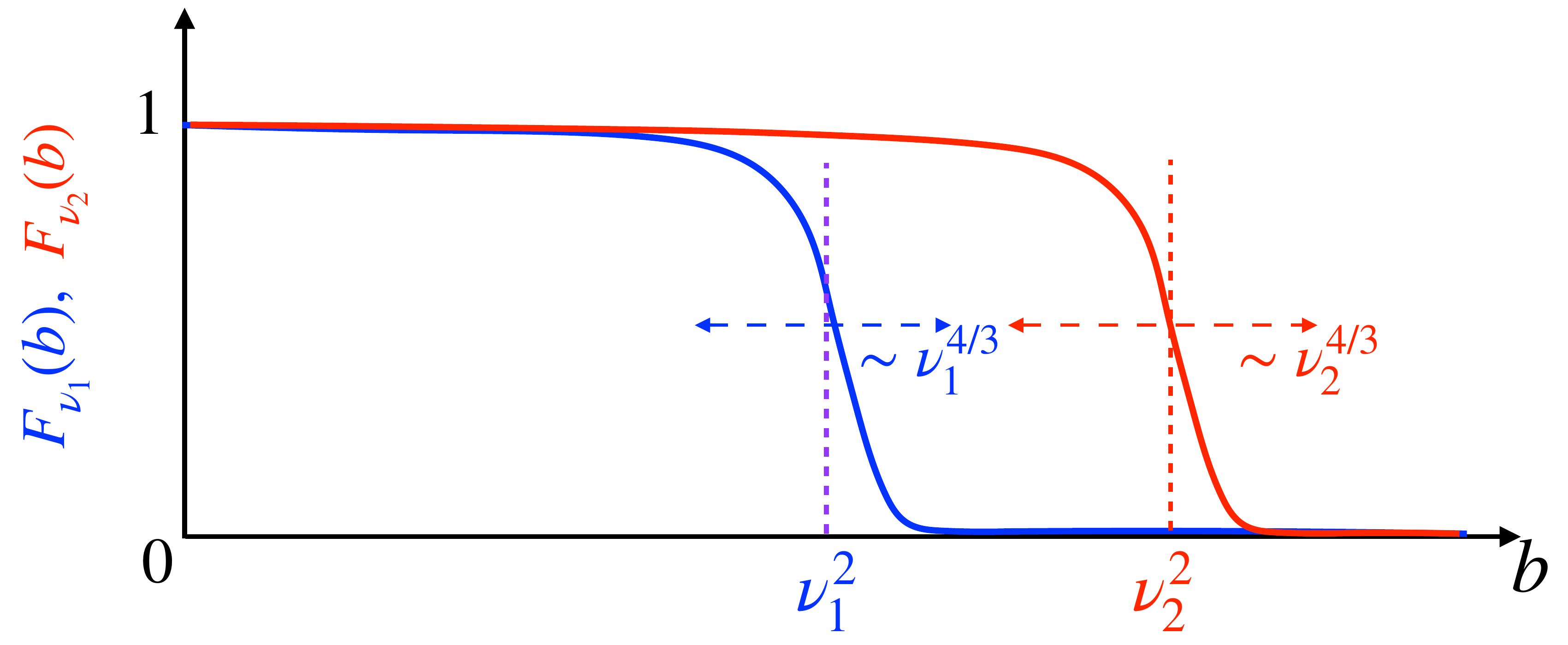}
    \caption{Schematic plot of $F_{\nu}(b)$ vs $b$ for two different large values of $\nu$, namely $1 \ll \nu_1<\nu_2$. The transition for $1$ to $0$ happens for $b \sim \nu^2$ on a scale $\propto \nu^{4/3}$.}
    \label{fig:Fnu}
\end{figure}
For integer $\nu$ (which corresponds to even space dimension for the fermions) there are other representations for the hole probability of the Bessel process from
the paper of Forrester and Hughes \cite{FH1994}
\bea \label{expr_FH}
F_\nu(b) = e^{- b/4} \det_{1 \leq j,k \leq \nu} I_{j-k}(\sqrt{b})  \;.
\eea 
The determinant in (\ref{expr_FH}) is a Toeplitz determinant which, using the Cauchy-Binet formula can be written as a matrix integral over the unitary group 
\bea \label{GW}
F_\nu(b)= e^{- b/4}  \int_{U(\nu)} e^{\frac{1}{2} \sqrt{b} {\rm Tr} (U + U^+)} \;.
\eea 

This matrix integral is well known in lattice quantum chromodynamics and was studied in the early eighties by Gross and Witten \cite{GW1980} and independently by Wadia \cite{Wadia}. More recently, this matrix integral appeared in the context of the longest increasing subsequence of a random permutation \cite{Joh1998,BDJ1999,BorodinForrester,Rains1998}. Consider the unit square filled by
points drawn from a Poisson point process of intensity $t$, and denote $L(t)$ the length of the optimal up/right path from
$(0,0)$ to $(1,1)$ which collects the maximum number of points (which is $L(t)$). Then one has ${\rm Prob}(L(t) < \nu) = F_\nu(b=4 t)$. 

Consider again the hole probability for the fermions in $d$ dimensions in the microscopic regime, which, as shown in the text, takes the form of the infinite product
\be \label{BesselProduct2} 
{\sf P}_d(z) =  \prod_{\ell=0}^{+\infty}  F_{\ell+ \frac{d}{2}-1} (b)^{g_d(\ell)}  \quad , \quad b = z^2
\ee 
where $z=\sqrt{2\mu} R$. We are interested now in the large $z$ behavior of ${\sf P}_d(z)$, i.e., we need
to consider the decay from $1$ to $0$ of each factor $F_\nu(b)$ at large $b$. As shown on the Figure \ref{fig:Fnu}, this decay occurs on the scale $b \sim \nu^2$, i.e., 
one has $F_\nu(b) \approx 1$ for $b \ll \nu^2$ and $F_\nu(b) \approx 0$ for $b \ll \nu^2$. Hence we need to consider large values of $\nu$.
%Indeed, as shown on the Figure \ref{fig:Fnu}, one has $F_\nu(b) \approx 1$ for $b \ll \nu^2$ and $F_\nu(b) \approx 0$ for $b \ll \nu^2$. 

Let us recall that $F_\nu(b)={\rm Prob}( b_{\min} > b)$ is the cumulative distribution function (CDF) of the (scaled) 
smallest eigenvalue of the WL ensemble. Its transition from $1$ to $0$ corresponds to an effective edge
which occurs in the large $\nu$ limit in the spectral density of the WL ensemble at location $\simeq \nu^2$. 
In Fig. \ref{Fig_TW} we have plotted the associated PDF $- F'_\nu(b)$ which exhibits three regimes characteristic
of a third order transition \cite{SM2014}.
Let us define the parameter $\gamma= \sqrt{b}/\nu$ so that the transition occurs at $\gamma=1$. 
In terms of $F_\nu(b)$ these regimes are
\\

(i) for $\gamma^2= b/\nu^2 < 1$, it corresponds to a pulled Coulomb gas, and exhibits the
large deviation form 
\be \label{regime1} 
F_\nu(b) \sim 1 - \exp\left( - \nu \phi_-(\gamma= \frac{\sqrt{b}}{\nu} ) \right) 
\ee
where the rate function $\phi_-(\gamma)$ is given by \cite{Se1998,KC2010} 
\bea \label{phi_m2}
\phi_-(\gamma) = 2 \left(\log \left(\sqrt{1-\gamma ^2}+1\right)-\log \gamma
  - \sqrt{1-\gamma ^2}  \right) \quad , \quad \gamma < 1 
\eea
and behaves as $\phi_-(\gamma) \simeq \frac{4 \sqrt{2}}{3}   (1-\gamma)^{3/2}$
for $\gamma \approx 1^-$ near the transition. Note that the log divergence of $\phi_-(\gamma)$ at small $\gamma$ leads to
$F_\nu(b) \sim 1 -  (\frac{b}{ \nu^2})^{\nu + o(\nu)}$, which matches the exact behavior
$1-F_\nu(b) \simeq b^{1 + \nu}$.

\vspace*{0.5cm}
\noindent(ii) for $\gamma^2= b/\nu^2 > 1$, it corresponds to a pushed Coulomb gas, and exhibits the
large deviation form
\be \label{Fnupush}
F_\nu(b) \sim \exp( - \nu^2 \phi_+(\gamma= \frac{\sqrt{b}}{\nu} ) ) 
\ee 
where the rate function $\phi_+(\gamma)$ is given by
\be \label{phi+}
\phi_+(\gamma) =   \frac{\gamma^2}{4} -\gamma + \frac{1}{2} \log \gamma + \frac{3}{4} \quad, \quad \gamma > 1 
\ee
and behaves as $\phi(\gamma) \simeq \frac{1}{6}   (\gamma-1)^3$ 
for $\gamma \approx 1^+$ near the transition.
This was obtained by Gross and Witten \cite{GW1980}, Wadia \cite{Wadia} and Johansson \cite{Joh1998},
using the Coulomb gas for the unitary matrix model (see also \cite{KC2010}). 

(iii) for $\gamma^2= b/\nu^2 = 1$. This is the critical regime. Indeed it is known that in the large index $\nu$ limit, the Bessel process takes the form
$b_i \simeq \nu^2 - 2^{2/3} \nu^{4/3} a_i$ where the $a_i=O(1)$ form the so-called Airy$_2$ determinantal point process (associated to the Airy kernel). Consequently, in this regime, $F_\nu(b)$ takes the scaling form \cite{BFP1998,BDJ1999}
\bea
F_\nu(b) \approx {\cal F}_2\left(\frac{\nu^2 - b}{2^{2/3} \nu^{4/3}}\right) \;,
\eea 
where ${\cal F}_2(z)$ is the $\beta = 2$ Tracy-Widom distribution, which describes the (scaled) fluctuations of the largest eigenvalue in the Gaussian Unitary Ensemble \cite{TW1994a}.

Consider now again the hole probability in \eqref{BesselProduct2} and
take its logarithm given by the infinite sum
\be \label{BesselProduct3} 
\log {\sf P}_d(z=\sqrt{b}) = \sum_{\nu=\frac{d}{2}-1}^{+\infty}  g_d(\nu-\frac{d}{2}+1)
\log F_{\nu} (b) = S_1 + S_2 + S_3
\ee
The $S_i$'s denote the contributions to the total sum coming from the three regimes (i-iii) above.
In each contribution we can approximate the sums over $\nu$ by an integral (neglecting also
$\frac{d}{2}-1$ compared to $\nu$) and use the asymptotics $g_d(\ell) \simeq \frac{2}{\Gamma(d-1)} \ell^{d-2}$. It is easy to see that the third term $S_3$ corresponding to regime (iii) is the dominant one. Indeed one has
\be 
S_1 \approx - \frac{2}{\Gamma(d-1)} \int_{\sqrt{b}}^{+\infty}  d\nu \nu^{d-2} 
\exp\left( - \nu \phi_-(\gamma= \frac{\sqrt{b}}{\nu} ) \right)
\propto (\sqrt{b})^{d-1} 
\int_{0}^{1}  d\gamma \gamma^{-d} 
\exp\left( - \frac{\sqrt{b}}{\gamma} \phi_-(\gamma)  \right)
\ee 
where we have used the logarithm of the form \eqref{regime1} and expanded $\log(1-z) \simeq -z$ 
for small $z$. In the last integral we have changed the integration variable from $\nu$ to $\gamma=\sqrt{b}{\nu}$. It is easy to see that this integral is dominated by the vicinity
of $\gamma=1$ where we can replace $\phi_-(\gamma) \simeq \frac{4 \sqrt{2}}{3}   (1-\gamma)^{3/2}$.
This approximation leads to an integral $\propto (\sqrt{b})^{-2/3}$ and finally
$S_1 \sim (\sqrt{b})^{d-\frac{5}{3}}$ which is negligible compared to the result
obtained below for $S_3$. A similar analysis shows that any finite window in the region (ii) 
gives a contribution of $O((\sqrt{b})^{d-\frac{4}{3}})$. The contribution from regime (iii)
is evaluated by taking the logarithm of the form in \eqref{Fnupush} leading to
\be 
S_3 \simeq - \frac{2}{\Gamma(d-1)} \int_0^{\sqrt{b}}  d\nu \nu^{d}
\phi_+(\gamma= \frac{\sqrt{b}}{\nu} ) 
\simeq (\sqrt{b})^{d+1} 
\int_{1}^{+\infty}  d\gamma \gamma^{-(d+2)} \phi_+(\gamma) 
\ee 
upon setting $b=z^2$ it coincides with formula \eqref{Pdl} in the text (recalling 
$\nu \simeq \ell$) where the integral over $\gamma$ is evaluated explicitly.
\\
\begin{figure}[t]
\includegraphics[width = 0.7 \linewidth]{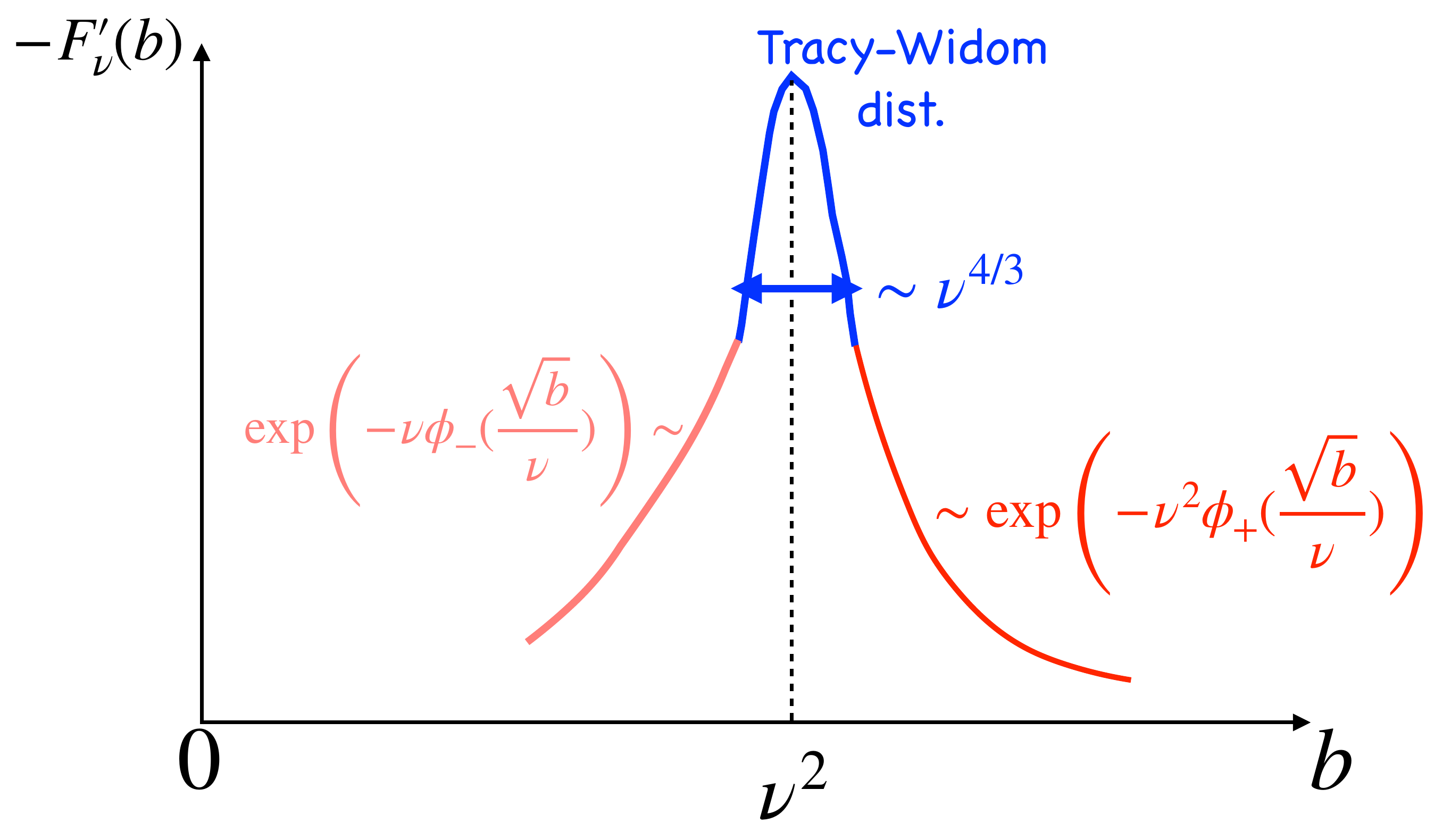}
\caption{Sketch of the PDF $-F'_{\nu}(b)$ of $b_{\min}$ in Eq. (\ref{PFD}) for $\nu \gg 1$, as a function of $b$ around the "edge" at $b \nu^2$. The regime of the typical fluctuations of $b_{\min}$ has a width $\sim \nu^{4/3}$ and is described by the 
Tracy-Widom distribution. The right and left tails correspond to large deviations regimes described by the rate functions $\phi_-(\gamma)$ in Eq. (\ref{phi_m2}) -- pulled Coulomb gas -- and $\phi_+(\gamma)$ in Eq. (\ref{phi_p}) -- pushed Coulomb gas.}\label{Fig_TW}
\end{figure}
{\it Derivation of the rate functions using Painlev\'e equation}.
It is instructive to give here also an alternative derivation of the formula for
the rate functions $\phi_+(\gamma)$ and $\phi_-(\gamma)$ from the Painlev\'e equation
which turns out to be valid for any $d$ (while the derivations in the references mentioned 
above \cite{GW1980,Wadia,Joh1998}
were performed for integer $\nu$). 
As mentioned in the text, the FD in \eqref{PFD} can be expressed as
an integral
%\log F_\nu(b)= - \int_0^b \frac{ds\, \sigma(s)}{s}$ 
from the solution
$\sigma(s)$ of the Painlev\'e III equation
\be \label{P3}
\log F_\nu(b)= - \int_0^b \frac{ds\, \sigma(s)}{s} \quad , \quad 
(s \sigma'')^2 + \sigma' (\sigma - s \sigma')(4 \sigma'-1) - \nu^2 (\sigma')^2 = 0  
\ee
where $\sigma(s) \simeq \frac{s^{1+\nu}}{2^{2 \nu+2} \Gamma(1+\nu) \Gamma(2+\nu)}$
at small $s$.

{\it Regime (ii)}. We are a looking for a form $\log F_\nu(b) \simeq - \nu^2 \phi_+(\frac{\sqrt{b}}{\nu})$. Taking a derivative w.r.t. $b$ we see that 
it corresponds to $\sigma(s)$ of the following form in the large $\nu$, large $s$ limit
with $s \sim \nu^2$
\be \label{scaling_sigma}
\sigma(s)= \nu \frac{\sqrt{s}}{2} f(\gamma= \frac{\sqrt{s}}{\nu} ) \quad , \quad f(\gamma)= \phi'_+(\gamma) 
\ee
Inserting this scaling form in the Painlev\'e
equation \eqref{P3} and neglecting subleading terms for large $\nu$ we obtain a differential equation for $f(\gamma)$
\bea \label{P3bis}
\gamma  f'(\gamma )+f(\gamma ) = 0 \quad \text{or} \quad
\gamma  f'(\gamma ) \left(-\gamma
   ^2+\gamma ^2 f'(\gamma )+1\right)+\left(\gamma ^2+1\right) f(\gamma )-\gamma 
   f(\gamma )^2 = 0 \;.
\eea
One can check that 
\be
f(\gamma) = \frac{\gamma }{2}+\frac{1}{2 \gamma }-1 
\ee 
is indeed solution of equation \eqref{P3bis}. Identifying 
$ \phi_+'(\gamma)=f(\gamma)$ and integrating, using the boundary condition
$\phi_+(1)=0$ we obtain the formula \eqref{phi+}
for $\gamma>1$.

{\it Regime (i)}. We are a looking for the form \eqref{regime1} i.e.,  
\bea
 \log F_\nu(b)  \sim - \exp( - \nu \phi_-(\gamma= \frac{\sqrt{b}}{\nu} ) ) 
\eea
Taking a derivative w.r.t. $b$ and neglecting the prefactor of the exponential we can search for a solution
of the Painlev\'e equation of the form
\be 
\sigma(s) \sim \exp( - \nu \phi_-(\gamma= \frac{\sqrt{s}}{\nu} ) + o(\nu)  ) 
\ee 
Inserting this form in \eqref{P3} and keeping the leading order at large $\nu$ we obtain that 
$\phi_-(\gamma)$ must satisfies $\phi_-'(\gamma)^2 (4 - \frac{4}{\gamma^2} + \phi_-'(\gamma)^2)=0$
which leads to the solution \eqref{phi_m2} for $0<\gamma<1$.  \\

\section{Numerical evaluation of the free fermion hole probability}

To evaluate numerically the free fermion hole probability ${\sf P}_d(z)$ we need to evaluate the product of Fredholm determinants in Eq. \eqref{BesselProduct2} associated to the Bessel process where 
$F_\nu(b=z^2)$ is given in \eqref{PFD}. This was performed using the Bornemann's method, an algorithm that allows to compute such determinant in an efficient manner
\cite{bornemann2010numerical}.
%Then we need to compute the Infinite product of Fredholm determinants. 
At this point one has to tune parameters on which the precision of the result depends. We made sure that those parameters ensure high enough precision by running multiple numerical tests. 
Because we cannot compute an infinite number of determinants, we use the fact that for a given $z$, high enough $\ell$ does not influence the total product. In practice, we cut the product at  $\ell = \ell_{\max}(z)={\rm Int}(1.1 \sqrt{z})$ and checked that the product converges when $\ell_{\max}(z)$ goes from 0 to ${\rm Int}(1.1 \sqrt{z})$. 
For $z<16$ we just set $\ell_{\max}(z)=4$.
%in order to avoid border effect when $\ell_{max}(z)$ becomes too small.

\section{Small $R$ expansion of the $d$ dimensional hole probability}

Here we perform the small $R$ expansion of the hole probability for the free fermions in $d$ dimension, $P(R)= {\sf P}(z= f_F R)$ defined in the text. 

\subsection{Using the $d$ dimensional kernel}

To obtain series expansions at small $z$ for ${\sf P}_d(z)$ we use the determinantal formula \eqref{Besselddim} in the text,
which we expand as an infinite series in powers of traces 
\be
\log {\sf P}_d(z) = \log {\rm Det}(I - P_z K_d) = {\rm Tr} \log (I - P_z K_d) =  - \sum_{k=1}^{+\infty} \frac{1}{k} {\rm Tr} (P_z K_d)^k
\ee 
where $P_z$ is the projector on the $d$-dimensional ball of radius $z$ and $K_d$ is the $d$ dimensional kernel for free fermions~\cite{Torquato2008,DeanPLDReview}
\be
K_d({\bf x}, {\bf x}') = \frac{J_{d/2}(|{\bf x}- {\bf x}'|) }{(2 \pi |{\bf x}- {\bf x}'|)^{d/2} }
\ee 
Upon rescaling variables ${\bf x} \to z {\bf x}$ we can perform all the integrals on the unit ball and write
\be \label{sumk} 
\log {\sf P}_d(z) =   - \sum_{k=1}^{+\infty} \frac{z^{k d}}{k} \, ( \prod_{j=1}^k \int_{{\cal S}_d} d^d{\bf x}_j ) \, 
\tilde K^d_z({\bf x}_1,{\bf x}_2) \dots \tilde K^d_z({\bf x}_k,{\bf x}_1) 
\ee 
where ${\cal S}_d$ is the $d$-dimensional sphere of unit radius. We have introduced the scaled kernel $\tilde K_z^d ({\bf x}, {\bf x}')$
which itself can be expanded in powers of $z$ as follows
\bea
&& \tilde K_z^d ({\bf x}, {\bf x}') = \frac{J_{d/2}(z |{\bf x}- {\bf x}'|) }{(2 \pi z |{\bf x}- {\bf x}'|)^{d/2} } \\
&& = A_d
\left( 1-\frac{z^2}{2 (d+2)} |{\bf x}- {\bf x}'|^2 +\frac{z^4}{8 (d+2)(d+4)} |{\bf x}- {\bf x}'|^4 +O(z^5) \right)
\quad , \quad A_d = \frac{1}{(2 \pi)^{d/2}} \frac{1}{2^{d/2} \Gamma(1+ \frac{d}{2})} \label{Kdz} 
\eea 
To each order $k$ in \eqref{sumk}, keeping only the leading order in \eqref{Kdz} we obtain 
\bea
-  \frac{z^{k d}}{k} \left[ \frac{S_d}{d} A_d  \right]^k =
-  \frac{z^{k d}}{k}  (B_d)^k  \quad , \quad     B_d=\frac{1}{2^d \Gamma(1+ \frac{d}{2})^2}
\eea
where $S_d= 2 \pi^{d/2}/\Gamma(d/2)$ is the area of the unit sphere and $S_d/d$ its volume.
For the term $k=1$ in \eqref{sumk} this is the exact result since $\tilde K^d_z$ is evaluated at coinciding points.

The next to leading term $O(z^2)$ in \eqref{Kdz} is given for $k \geq 2$ by
\bea
&&   \frac{z^{2+k d}}{2 (d+2)} A_d^k \left[ \frac{S_d}{d}   \right]^{k-2} S_d S_{d-1} \int_0^1 dx_1 x_1^{d-1} \int_0^1 dx_2 x_2^{d-1} 
  \int_0^{\pi} d\theta (\sin \theta)^{d-2} 
  (x_1^2 - 2 x_1 x_2 \cos(\theta) + x_2^2) \\
% && =  \frac{z^{2+k d}}{2 (d+2)}  \left[ A_d \frac{S_d}{d} \right]^{k} \frac{S_d S_{d-1}}{(S_d/d)^2}   \frac{\sqrt{\pi} \Gamma(\frac{d-1}{2})}{2 \Gamma(2+ \frac{d}{2})} \\
 && = \frac{z^{2+k d}}{2 (d+2)} B_d^{k} \frac{S_d S_{d-1}}{(S_d/d)^2}   \frac{\sqrt{\pi} \Gamma(\frac{d-1}{2})}{2 \Gamma(2+ \frac{d}{2})} 
 = B_d^{k} \frac{d}{(d+2)^2} z^{2+k d}
\eea
To calculate the next order for $k=2$ we first include the next subleading term $O(z^4)$ in \eqref{Kdz}
\bea
&& - \frac{z^{4+k d}}{8 (d+2)(d+4)}
 A_d^k \left[ \frac{S_d}{d}   \right]^{k-2} S_d S_{d-1} \int_0^1 dx_1 x_1^{d-1} \int_0^1 dx_2 x_2^{d-1} 
  \int_0^{\pi} d\theta (\sin \theta)^{d-2} 
  (x_1^2 - 2 x_1 x_2 \cos(\theta) + x_2^2)^2 \\
 && = - \frac{z^{4+k d}}{8 (d+2)(d+4)} B_d^{k} \frac{S_d S_{d-1}}{(S_d/d)^2}   \frac{(3+d)\sqrt{\pi} \Gamma(\frac{d-1}{2})}{2 \Gamma(3+ \frac{d}{2})} 
 = - B_d^{k} \frac{d (d+3)}{2 (d+2)^2 (d+4)^2} z^{4+k d}
\eea 
and then we also need to consider the square of \eqref{Kdz} which gives to that order
\bea
&& - \frac{1}{2} \frac{z^{4+k d}}{(2 (d+2))^2} B_d^{k} \frac{S_d S_{d-1}}{(S_d/d)^2}   \frac{(3+d)\sqrt{\pi} \Gamma(\frac{d-1}{2})}{2 \Gamma(3+ \frac{d}{2})} 
 = - B_d^{k} 
 \frac{d (d+3)}{2 (d+2)^3 (d+4)} z^{4+k d} 
\eea
Putting the two contributions together for $k=2$ we finally obtain, in any $d$
\bea \label{small_z_sm}
\log {\sf P}_d(z) =  - \sum_{k=1}^{+\infty} \frac{z^{k d}}{k} B_d^k 
+ \sum_{k=2}^{+\infty} B_d^{k} \frac{d}{(d+2)^2} z^{2+k d} - \frac{d (d+3)^2}{(d+2)^3 (d+4)^2} z^{4+2 d} B_d^2 + \dots
\eea
It can be rewritten as
\bea
P_d(z) =  (1 - z^d B_d) e^{ \frac{d}{(d+2)^2} \frac{z^{2+2 d} B_d^2}{(1- B_d z^d)} - \frac{d (d+3)^2}{(d+2)^3 (d+4)^2} z^{4+2 d} B_d^2 + \dots }
\eea
which, if we discard the subdominant term in the exponential yields the formula \eqref{small_z_intro} in the text. 
%It can also be stated as follows
%\bea
%&& \log P_d(z) =  - z^d B_d - z^{2 d} B_d^2 \left( \frac{1}{2} - \frac{d}{(d+2)^2} z^2 + \frac{d (d+3)^2}{(d+2)^3 (d+4)^2} z^4 + O(z^6) \right) \\
%&& - \sum_{k \geq 3}^{+\infty} B_d^{k} z^{k d} \left( \frac{1}{k} -  \frac{d}{(d+2)^2} z^{2} + O(z^4) \right)
%\eea
For $d=2,3,4$ this leads to
\bea
&& \log {\sf P}_2(z) = - \frac{1}{4} z^2 - \frac{1}{32} z^{4} + \frac{1}{384} z^6 - \frac{7}{18432} z^8 + O(z^{10})  \;,
\eea
leading to 
\bea
{\sf P}_2(z) = 1-\frac{z^2}{4}+\frac{z^6}{128}-\frac{25 z^8}{18432}+O\left(z^9\right) \;.
\eea

\bea
&& \log {\sf P}_3(z) =
-\frac{2 z^3}{9 \pi }-\frac{2 z^6}{81 \pi ^2}+\frac{4
   z^8}{675 \pi ^2}-\frac{8 z^9}{2187 \pi ^3}-\frac{16
   z^{10}}{18375 \pi ^2}+\frac{8 z^{11}}{6075 \pi
   ^3}-\frac{4 z^{12}}{6561 \pi ^4}+O\left(z^{13}\right) \;,
\eea
leading to
\bea
{\sf P}_3(z) = 1-\frac{2 z^3}{9 \pi }+\frac{4 z^8}{675 \pi ^2}-\frac{16
   z^{10}}{18375 \pi ^2}+O\left(z^{13}\right)
\eea

\bea \label{d4}
\log {\sf P}_4(z) = - \frac{1}{64}\,z^4 - \frac{1}{8192}\,z^8 + \frac{1}{36864}\,z^{10} - \frac{67}{14155776} z^{12} + {\cal O}(z^{14}) \;,
\eea 
leading to
\bea
{\sf P}_4(z) = 
1-\frac{z^4}{64}+\frac{z^{10}}{36864}-\frac{49 z^{12}}{14155776}+O\left(z^{13}\right) \;.
\eea
\\

Let us recall that in $d=1$, i.e., the hole probability for the interval $[-z,z]$ (of size $2z$), the above results lead to 
\bea
{\sf P}_1(z) = 
1-\frac{2 z}{\pi }+\frac{4 z^4}{9 \pi ^2}-\frac{64 z^6}{675 \pi ^2} + O\left(z^8\right) \;.
\eea 
This agrees with the results in \cite{Grimm2004} and \cite{DeanPLDReview} (integrating twice Eq. (46)
in \cite{DeanPLDReview} and setting $s=2 z/\pi$). 

{\bf Remark}. For even space dimension $d$, one can use formulae \eqref{BesselProduct} and \eqref{eq:forrester}. 
In practice to evaluate the expansion up to a given order in $z$, one needs only a finite number of terms corresponding to the lowest
values of $\ell$ in the product \eqref{BesselProduct}. Using Mathematica we have checked the above formula using this procedure for
$d=2,4$.

\subsection{Using the angular decomposition and the Painlev\'e equation}\label{sec:pain}

An alternative way to obtain the small $z$ expansion is to perform a
small $b$ expansion of each term $\log F_\nu(b)$ in \eqref{BesselProduct}
using the Painleve equation \eqref{P3}. We recall that $\nu \geq \frac{d}{2}-1$
is an integer (for even $d$) or half-integer (for even $d$).
One uses the small $s$ asymptotics of $\sigma(s)$ given in formula (1.22) \cite{TWB1994}
with $\alpha=\nu$. Here for $d \geq 2$ we need only, for $\nu \geq 1/2$
\be
\sigma(s) = \frac{1}{2^{2 \nu+2} \Gamma(1+\nu) \Gamma(2+\nu)} s^{1+\nu}(1 - \frac{s}{2 (2+\nu)} + O(s^2)) + O(s^{2+ 2 \nu}) 
\ee
Integrating once we obtain the asymptotics of $\log F_\nu(b)$ at small $b$ for $\nu \geq 1/2$ as
\bea \label{Fnularge} 
&& \log F_\nu(b) 
%= - \frac{1}{2^{2 \nu+2} \Gamma(2+\nu)^2} b^{1+\nu}(1 - \frac{1+\nu}{2 (2+\nu)^2} b 
%+ O(b^2)) + O(b^{2+ 2 \nu}) \\
=  - \frac{1}{2^{2 \nu+2} \Gamma(2+\nu)^2} b^{1+\nu} 
+ \frac{1+\nu}{2^{2 \nu+3} \Gamma(3+\nu)^2} b^{2+\nu} + O(b^{2+ 2 \nu})
\eea
Let us note that $\nu=0$ (which occurs only in $d=2$) is a special case for which the above formula does not apply and
\be
\sigma(s)=\frac{s}{4}   \quad , \quad \log F_0(b) = - \frac{b}{4} 
\ee

To evaluate the lowest orders in $z=\sqrt{b}$ in \eqref{BesselProduct} we write
\be
{\sf P}_d(z) = \exp( \log F_{\frac{d}{2}-1}(z^2) + g_d(1) \log F_{\frac{d}{2}}(z^2)  + g_d(2) \log F_{\frac{d}{2}+1}(z^2)  + \dots ) 
\ee
with $g_d(1)=d$, $g_d(2)=\frac{(d+2)(d-1)}{2}$.
We use that for $d \geq 3$
\be
\log F_{\frac{d}{2}-1}(z^2) = - \frac{1}{2^{d} \Gamma(1+ \frac{d}{2})^2} z^d (1 - \frac{d}{4 (1+ \frac{d}{2})^2} z^2 + O(z^4))  
% + \frac{1}{(1+ \ell) (1+ 2 \ell)} [\frac{1}{2^{2 \ell+2} \Gamma(1+\ell) \Gamma(2+\ell)}]^2 
+ \frac{2}{d(d-1)} [ \frac{1}{2^d \Gamma(\frac{d}{2}) \Gamma(\frac{d}{2} +1)} ]^2 z^{2 d} + \dots
\ee
and for $d=2$ one has 
\be
\log F_{\frac{d}{2}-1}(z^2) = - \frac{1}{4} z^2  
\ee
We also use that, for $d \geq 2$
\be
\log F_{\frac{d}{2}}(z^2) = - \frac{1}{2^{d+2} \Gamma(2+\frac{d}{2})^2} z^{2+d}(1 - \frac{1+\frac{d}{2}}{2 (2+\frac{d}{2})^2} z^2 + O(z^4)) + O(z^{4+ 2 d}) 
\ee
We also need only the leading order of
\be
\log F_{\frac{d}{2}+1}(z^2) = - \frac{1}{2^{4+d} \Gamma(3 + \frac{d}{2})^2} z^{4+d} 
\ee

Using these expansions, putting together all terms we recover the lowest orders of the same series expansions performed in the 
previous section in Eq. (\ref{small_z_sm}). Note that, within this method, many cancellations arise between the contributions of the various angular sectors.

\section{Exact formula for the hole probability for the harmonic potential} 

\subsection{General formula in even space dimension}

Let us give here the exact expression for the hole probability for a finite number 
$N$ fermions in the harmonic potential $V(r)=\frac{1}{2} r^2$ in even dimension $d$.

Let us recall how the harmonic potential is filled. Recall that the energy levels are $\epsilon_{n,\ell}=2 n + \ell + \frac{d}{2}$, with $n, \ell$ positive integers.
In even dimension $d$ one takes the Fermi energy $\mu$ to be an integer, and in odd dimension $d$ a half integer. 
The minimal value of $\mu$ which corresponds to a single fermion $N=1$ in the well is $\mu= \frac{d}{2}$ (corresponding to the occupied state $n=0$ and $\ell=0$).
Upon increasing $\mu$ by integer units, the total number of fermions in the well is given by $N = \sum_{n=0}^{+\infty} \sum_{\ell=0}^{+\infty} g_d(\ell) \theta(\epsilon_{n,\ell} \leq \mu)$,
which correspond to non degenerate ground states. We recall that $g_d(\ell)$ is the degeneracy of the eigenenergy $\epsilon_{n,l}$ and its expression is given by
\be \label{gd_l}
g_d(\ell) = \frac{(2 \ell + d-2) \Gamma(\ell+d-2)}{\Gamma(\ell+1) \Gamma(d-1)} \;, \; \ell \geq 1 \;.
\ee 
This can also be written as 
$N= \sum_{\ell=0}^{\ell_{\max}} g_d(\ell) m_l$, where $m_\ell = {\rm Int} \left(\frac{\mu-\ell-\frac{d}{2}}{2} + 1\right)$ is the number of fermions in the
angular sector $\ell$, and $\ell_{\rm max}(\mu)= \mu- \frac{d}{2}$. 

Let us give as an example the case of $d=2$. In that case one finds that for $\mu=2 k+1 $, $k \geq 0$, one has $N= (k+1)(2 k+1)$ and for $\mu=2 k$, $k \geq 1$, one has $N=k(2 k+1)$.
The successive non degenerate states thus correspond to the values of the couples $(\mu,N)$ given by $(1,1)$, $(2,3)$, $(3,6)$, $(4,10)$ and so on. 

The hole probability $P(R)$ defined in the text is then a product of $\ell_{\max}$ terms
\be
P(R) = \prod_{\ell=0}^{\ell_{\rm max}(\mu)} P_\ell(R)^{g_d(\ell)} 
\label{product2} 
\ee 
where $P_\ell(R)$ is the hole probability in a given angular sector $\ell$ (i.e., the probability that there is no fermion in the 
interval $[0,R]$ in one dimension. We now use the correspondence between the positions of the fermions and 
the eigenvalues of the WL ensemble defined in \eqref{WL}. As explained in the text 
$P_\ell(R)={\rm Prob}(r_{\rm \min}>R)={\rm Prob}(\lambda_{\min} > R^2)$. In even dimension $d$, we can use the result (3.19) in
\cite{FH1994} and obtain $P_\ell(R)$ as a $\nu \times \nu$ determinant, with $\nu= a = \ell+ \frac{d}{2}-1$, given by
\be \label{HOhole} 
P_\ell(R)  = e^{-m_\ell R^2} \left| {\rm det}_{1 \leq j , k \leq \nu} \left[\left(\frac{d}{dt}\right)^{j+k-2} L_{\nu + m_\ell-1}^{1-\nu}(t) \right]_{t=-R^2}\right| \;.
\ee
where the $L_n^\alpha(x)$ are the generalized Laguerre polynomials. 
The equations \eqref{product2} and \eqref{HOhole} are thus an exact formula for the hole probability $P(R)$ for a finite 
number of fermions (where $N$ belongs to the sequence of integers described above).

%\begin{align*}
%&m \rightarrow m_{\ell}=\frac{\mu +1-(\ell+d/2-1)}{2} \\
%&n \rightarrow m_{\ell}+\ell+d/2-1
%\end{align*}

\subsection{Large and small $R$ asymptotics}

{\bf Small $R$ behavior}. To obtain the small $R$ asymptotics of $P(R)$ we need to first extract the small $R$ behavior of each $P_\ell(R)$ in \eqref{product2}. 
For this we consider the joint PDF in \eqref{WL} 
\be  \label{WL2} 
P^{(m)}_{\rm WL}(\vec \lambda) = \frac{1}{Z(m,\nu)} e^{- \sum_{i=1}^m \lambda_i} \prod_{i=1}^m \lambda_i^{\nu} 
\prod_{1 \leq j,k \leq m} (\lambda_j -\lambda_k)^2
\ee
where $m=m_\ell$ and we now keep track of the normalization, which is given by
\be 
Z(m,\nu) = m ! \prod_{j=0}^{m-1} \Gamma(j+1) \Gamma(\nu + j +1) 
\ee 
It is easy to extract the leading term using that
\bea
- \frac{d}{dR}  P_\ell(R)= - \frac{d}{dR} {\rm Prob }(\lambda_{\min}>R^2 ) = 2 m_\ell R^{2 \nu+1} \frac{Z(m_\ell-1,\nu+2)}{Z(m_\ell,\nu)} + o(R^{2 \nu+1}) 
\eea
where we have set one of the $\lambda_i$ to $R^2$ and approximated the remaining integrals by setting $R=0$. After some algebra this yields
%{\red P: Gabriel can you check that formula with your mathematica programs?} 
\be  \label{expansion1} 
P_\ell(R) = 1 - \frac{\Gamma(1+m+\nu)}{\Gamma(\nu+2)^2 \Gamma(m) } R^{2 \nu+2} + o(R^{2 \nu+2}) 
\ee 
where we recall that $\nu=a=\ell+\frac{d}{2}-1$ and $d$ even. We note that in the limit of large $m_\ell$, this formula matches the 
one in \eqref{Fnularge} for the free fermion/Bessel kernel regime with $b=z^2$ and $z=R k_F=R\sqrt{2 \mu}$. We have used that $m_\ell \simeq \mu/2$
in that regime.

Inserting \eqref{expansion1} into \eqref{product2} we see that the leading small $R$ behavior of the hole probability $P(R)$ is dominated by the term with smallest value of $\nu$,
i.e., $\ell=0$. Using that $g_d(0)=1$ we finally obtain 
\be 
P(R) = 1 - \frac{\Gamma(m_0+ \frac{d}{2})}{\Gamma(m_0) \Gamma(1 + \frac{d}{2})^2} R^d + o(R^d)
\ee 
where $m_0= {\rm Int}(\frac{\mu + 2 - \frac{d}{2}}{2})$. This is valid for even space dimension, and arbitrary integer value of $\mu$ (where $N$ is related to $\mu$ as
explained above). 
In the limit of large $\mu$ and $N$ this formula matches the universal result \eqref{small_z_intro}. 
\\
\begin{figure}[t]
\includegraphics[width=0.7\linewidth]{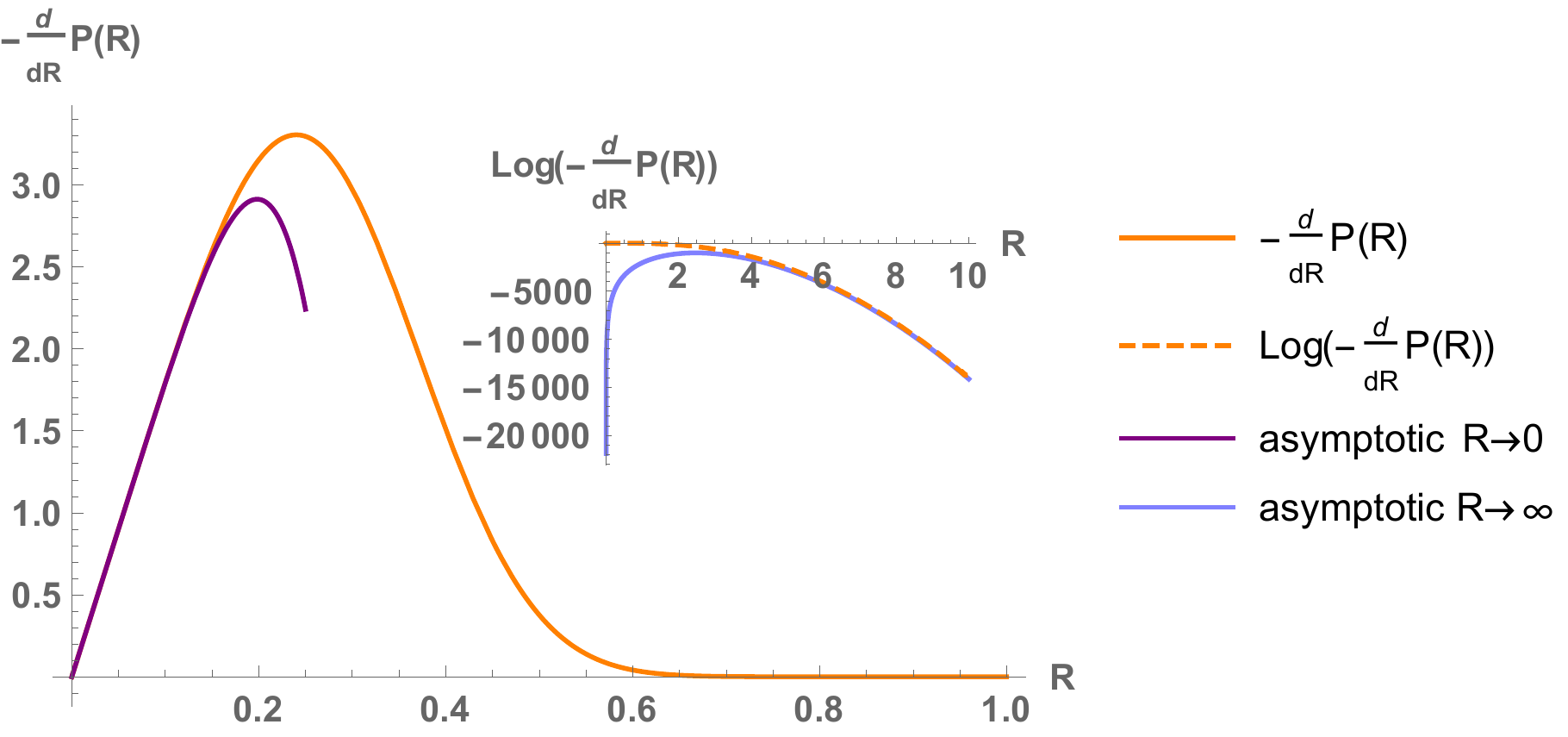}
\caption{$-\frac{d}{dR}P(R)$ vs. $R$ in dimension 2 for $N=171$ fermions in a harmonic trap $V(r)=\frac{1}{2} r^2$.
We have computed small $R$ (purple) and large $R$ (blue) behavior, the later being shown in the inset in semi-logarithmic scale. 
The large $R$ behavior has the general form given in Eq. \eqref{harmoniclargeR}. Both fit well the curve. Note that the typical scale in $R$
is of the order $1/\sqrt{2 \mu}$ with $\mu=18$.}
\label{fig:finiteN}
\end{figure}

{\bf Large $R$ behavior}. Now we want to extract the large $R$ behavior of $P_\ell(R)$. This is done by keeping only the largest coefficient of the polynomials inside the
determinant in \eqref{HOhole}. Using that $L_n^\alpha(x) \sim (-x)^n/n!$ at large $x$, we obtain the leading order of the derivatives of the Laguerre polynomials as
\be
\left(\frac{d}{dt}\right)^{j+k-2} L_{\nu + m_\ell-1}^{1-\nu}(t)\rightarrow (-1) ^{\nu+m_\ell - 1}f_{m_{\ell}+\nu+1-j-k}(t)
\ee
with $f_p(t)=\left\{
    \begin{array}{ll}
       \frac{t^p}{p!} & \mbox{if } p\geq 0 \\
        0 & \mbox p<0
    \end{array}
\right.$. Note that this can be zero if the Laguerre polynomial has been derived too many times.
Inserting into the determinant we obtain the estimate
\be
 \left| {\rm det}_{1 \leq j , k \leq \nu} \left[\left(\frac{d}{dt}\right)^{j+k-2} L_{\nu + m_\ell-1}^{1-\nu}(t) \right]_{t=-R^2}\right| \; \simeq
 \begin{vmatrix} 
   f_{m_{\ell}+\nu-1}(R^2) & f_{m_{\ell}+\nu-2}(R^2) & ... & &  f_{m_{\ell}}(R^2)  \\
   f_{m_{\ell}+\nu-2}(R^2) & f_{m_{\ell}+\nu-3}(R^2) && & . \\
  . & & & & .  \\
  .& & & & \\
  f_{m_{\ell}}(R^2)  &  & & & f_{m_{\ell}-\nu+1}(R^2) \\
   \end{vmatrix} 
\ee
Since $m_{\ell}$ is always positive, all the terms above the anti diagonal are guaranteed to be non zero, whereas the terms below can be zero depending on the value of $\ell$. 
From there we can compute each term of the Leibniz formula which all have the same degree.
%$(R^2)^{(\ell+d/2-1)m_{\ell}}$. 
For example it can be obtained by computing the product on the anti diagonal. %
This leads to the large $R$ asymptotics
%or the diagonal terms, but also for any more complicated permutation. Then we can factorize the $(R^2)^{(\ell+d/2-1)m_{\ell}}$ power and get:
\begin{align}
% \left| {\rm det}_{1 \leq j , k \leq \nu} \left[\left(\frac{d}{dt}\right)^{j+k-2} L_{\nu + m_\ell-1}^{1-\nu}(t) \right]_{t=-R^2}\right| \; \rightarrow 
 P_\ell(R) \simeq c_{\ell}(\mu)(R^2)^{(\ell+d/2-1)m_{\ell}} e^{-m_\ell R^2} \quad , \quad 
 c_{\ell}(\mu)=\begin{vmatrix} 
   f_{m_{\ell}+\nu-1}(1) & f_{m_{\ell}+\nu-2}(1) & ... & &  f_{m_{\ell}}(1)  \\
   f_{m_{\ell}+\nu-2}(R^2) & f_{m_{\ell}+\nu-3}(1) && & . \\
  . & & & & .  \\
  .& & & & \\
  f_{m_{\ell}}(1)  &  & & & f_{m_{\ell}-\nu+1}(1) \\
   \end{vmatrix} 
\end{align}
Here we will not attempt to compute the amplitude $c_\ell(\mu)$ (we checked on a few cases that it does not vanish).
Inserting this result in \eqref{HOhole} one found the following asymptotic formula for the hole probability
\begin{align}\label{harmoniclargeR}
P(R)&\underset{R \to \infty}{\sim} (\prod_{\ell\ge 0}^{\mu-d/2}c_{\ell}(\mu)^{g_d(\ell)})e^{-(\sum_{\ell\ge 0}^{\mu-d/2} g_d(\ell)m_{\ell}) R^2} R^{2 \sum_{\ell\ge 0}^{\mu-d/2} 
g_d(\ell)(\ell+d/2-1)m_{\ell}} \nonumber \\
&\underset{R \to \infty}{\sim} C(\mu)e^{-N R^2} R^{2M} 
\end{align}
where 
$N$ is the total number of fermions and $M=\sum_{\ell\ge 0}^{int(\mu-d/2)} g_d(\ell)(\ell+d/2-1)m_{\ell}$.
\\

In Fig. \ref{fig:finiteN} we show a plot of (minus) the derivative of the hole probability $-P'(R)$ in $d=2$ and for $\mu=18$ that is $N=171$.
We also show the asymptotic behaviors for small $R$ 
and for large $R$ in \eqref{harmoniclargeR} with the exact value of $C(\mu)$ evaluated with Mathematica.

\subsection{Large $N$ limit and free fermions}
\begin{figure}[t]
\includegraphics[width=0.6\linewidth]{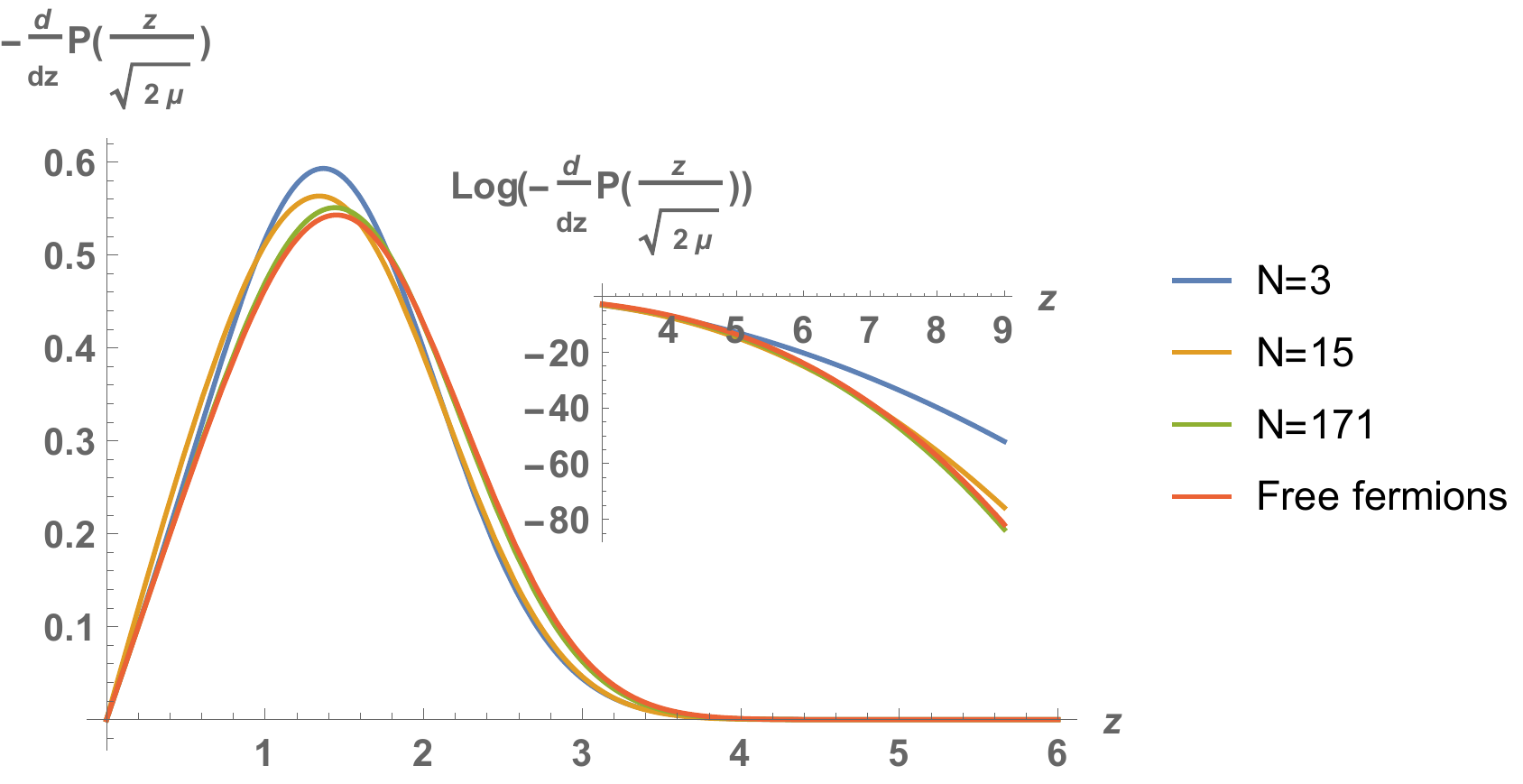}
\caption{Plot of the rescaled hole probability $-\frac{d}{dz}P(z/\sqrt{2 \mu})$ vs. $z$ for an harmonic trap in two dimensions $d=2$, for various values of $(\mu,N)$: $(2,3)$, $(5,15)$ and
$(18,171)$. We have also plotted the result for the free fermions, obtained from the Fredholm determinant formula using the kernel of the Bessel process, as
was computed in Fig. \ref{fig:freefermions}. Large $z$ behavior is shown in the inset as a semi-logarithmic scale.}
\label{fig:limitshape}
\end{figure}
Let us study how the result \eqref{HOhole} for $P(R)$ depends on the number of fermions. In particular we expect that in the large $N$ limit, i.e., the large $\mu$ limit,
the hole probability takes the scaling form $P(R) \simeq {\sf P}_d(z=\sqrt{2 \mu} R)$ where ${\sf P}_d(z)$ was obtained in the text and is also the result for free fermions.
This is studied in the Fig. \ref{fig:limitshape} where we show how the finite $N$ probability converges towards the free fermions limit shape.
In the bulk of the distribution the convergence in $N$ is fast. In the tail, which is shown in the inset, the convergence is slower.
More precisely, we observe that for $z=R\sqrt{2 \mu}<4$, the curves converge rapidly. 
For larger $z$, it fits to the free fermions limit shape up to a given value of $z$ which grows as $N$ increases.

%{\it Note on the numerical evaluation of the free fermion hole probability}.
%The Infinite product of Fredholm determinants was evaluated numerically. 
%Because we can not compute infinite number of determinants, we use the fact that for a given $z$, 
%high enough $\ell$ does not influence the total product. We set $\ell_{max}(z)=1.1 \sqrt{z}$ 
%and checked that the product converges. For $z<16$ we just set $\ell_{max}(z)=4$

\section{Large deviations regime for the harmonic potential} 

\subsection{General large deviation form of the hole probability in dimension $d>1$}

In this section we give some details about the regime $R \sim \sqrt{2 \mu}$. As explained in the text
the leading contribution to the product form \eqref{product} of the hole probability comes from
the region (iii), i.e., $R^2 > \lambda_- = m_\ell \zeta_-$, equivalently $R > r_-(\ell)$,
where the single sector hole probability is given by the large deviation form
\be
P_{\ell}(R) \sim e^{- 2 m_\ell^2 \Phi_+(\frac{R^2- m_\ell \zeta_-}{m_\ell}, \frac{\ell}{m_\ell}) }
\ee
where $\Phi_+$ is given below explicitly.
Inserting this form in the product \eqref{product} we see that the product is dominated by large $\ell = O(\mu)$,
in which case $m_\ell = \frac{\mu - \ell}{2}$ and one must have $\ell <\mu$. The question is which values of $\ell$ belong to
the regime (iii). This is indicated in Fig. \ref{fig:rm} where we plot $r_-(\ell)^2$ as a function of $\ell$, for $\ell \in [0,\mu]$.
We see on this figure that if $R^2 > \mu$, then all values of $\ell \in [0,\mu]$ will contribute since
$R$ is always larger than $r_-(\ell)$.
On the other hand, if $R^2< \mu$ there exist a root to the equation $r_-(\ell) = R$, denoted 
$\ell_+(R) = R \sqrt{2 \mu-R^2}$. For $\ell < \ell_-(R)$ one has $R> r_-(\ell)$ and for 
$\ell > \ell_+(R)$ one has $R < r_-(\ell)$ and regime (iii) does not contribute. As a consequence, the product over $\ell$ is in effect 
truncated for $\ell \leq \ell_+(R)$. For convenience we will extend the definition $\ell_+(R)=\mu$ for $R>\sqrt{\mu}$.

Let us now perform the product \eqref{product} replacing the sum by an integral
using $g_d(\ell) \sim \frac{2 \ell^{d-2}}{\Gamma(d-1)}$. We obtain
\bea
&& P(R)  \sim
\exp \left( - \frac{1}{\Gamma(d-1)} \int_0^{\ell_+(R)}  d \ell \ell^{d-2} (\mu -\ell)^2  \Phi_+(\frac{2 R^2}{\mu-\ell} -  \zeta_-, 
\frac{2 \ell}{\mu-\ell}) \right) \label{HP1} 
\eea 
where we recall that $\zeta_-\equiv\zeta_-(\ell)= (1 - \sqrt{1 + \frac{\ell}{m_\ell}})^2$. From the above discussion, the first argument of
the function $\Phi_+$ is positive for all values of $\ell \in [0,\ell_+(R)]$, as required. 
\begin{figure}[t]
    \centering
    \includegraphics[width=0.5 \linewidth]{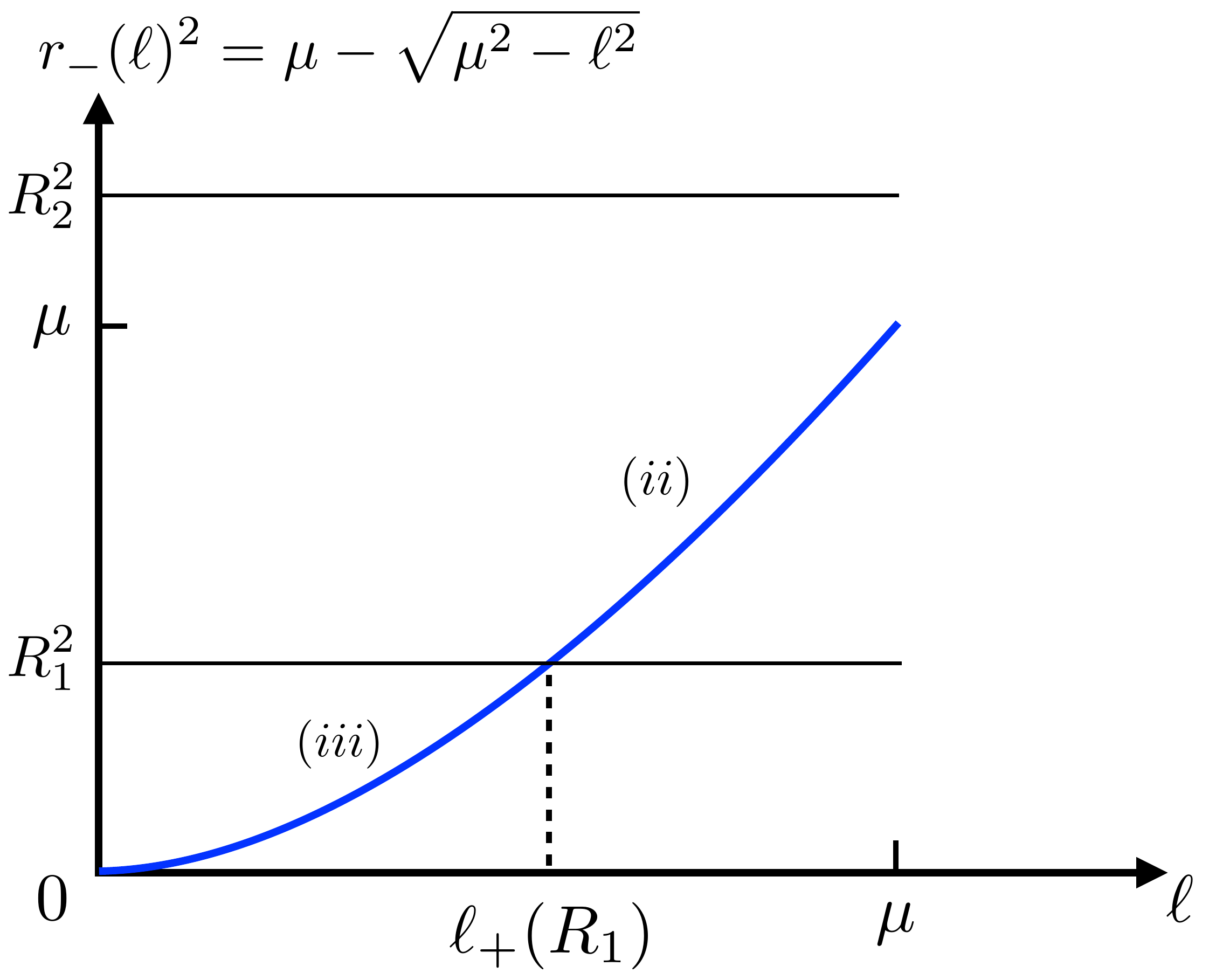}
    \caption{Sketch of the plot $r_-(\ell)^2$ vs $\ell$.}
    \label{fig:rm}
\end{figure}
Let us now scale all variables using $\mu$, and introduce $\tilde R=R/\sqrt{2 \mu}$ and the variable $v = \ell/\mu$ which varies from $0$ to $v_{\max}(\tilde R)$
with 
\bea 
v_{\max}(\tilde R) = 
\begin{cases}
&2 \tilde R \sqrt{1- \tilde R^2} \quad, \quad \tilde R^2 < 1/2 \\
&1 \quad, \quad \quad \quad \quad \quad \quad  \tilde R^2 > 1/2
\end{cases}
\eea 
We also introduce the variable $\alpha_v=2 v/(1-v)$. The formula \eqref{HP1} can then be rewritten as in the text
\bea
&& P(R) \sim \exp \left( - (2 \mu)^{d+1} \Psi(\tilde R) \right) 
\eea
in terms of the rate function
\bea \label{result2} 
&& \Psi(\tilde R) = 
\int_0^{v_{\rm max}(\tilde R)} dv\,\frac{v^{d-2} (1 - v)^2 }{2^{d+1} \Gamma(d-1)}  \Phi_+\left(\frac{4 \tilde R^2}{1-v} -  (1 - \sqrt{1 + \alpha_v})^2, 
\alpha_v\right) \;.
\eea

\subsection{Explicit expression for the rate function $\Psi(\tilde R)$ and its asymptotics}

In this subsection we study the explicit form of the rate function $\Psi(\tilde R)$ defined by the integral in \eqref{result2}. For this we recall the expression for the large deviation function of the WL ensemble in the pushed regime, as given in \cite{KC2010} (see also \cite{VivoMajumdar}). It is a function of two arguments, $x>0$, $\alpha>0$ given by
\bea
&& \Phi_+(x,\alpha) = \frac{1}{2} (S(x+ \zeta_-,\alpha) - S(\zeta_-,\alpha)) \\
&& S(\zeta,\alpha) = \frac{\zeta + U}{2} - \frac{(U-\zeta)^2}{32} - \log \frac{U-\zeta}{4}
+  \frac{\alpha}{4}  (\sqrt{U}-\sqrt{\zeta})^2 + \frac{\alpha^2}{4} \log (\zeta U) - \alpha(\alpha+2) 
\log \frac{\sqrt{U}+\sqrt{\zeta}}{2} \label{expr_S}\\
&& U = U(\zeta,\alpha)= \frac{4}{3} (\zeta + 2 (\alpha+2))   \cos^2( \frac{\theta + 2 \pi}{3} ) \quad , \quad 
%1 + \tan^2 \theta = \frac{1}{27 \alpha^2 \zeta} (\zeta + 2 (\alpha+2))^3 \quad , \quad 
\cos^2 \theta = \frac{27 \alpha^2 \zeta}{(\zeta + 2 (\alpha+2))^3} \nonumber
\eea 
with $\tan \theta>0$. Inserting this expression inside the formula \eqref{result2} leads a complicated expression for $\Psi(\tilde R)$.
The integral over $v$ can be performed numerically and the resulting function $\Psi(\tilde R)$ is plotted
in Fig. \ref{fig:largedev}. 

\begin{figure}[t]
\includegraphics[scale=.8]{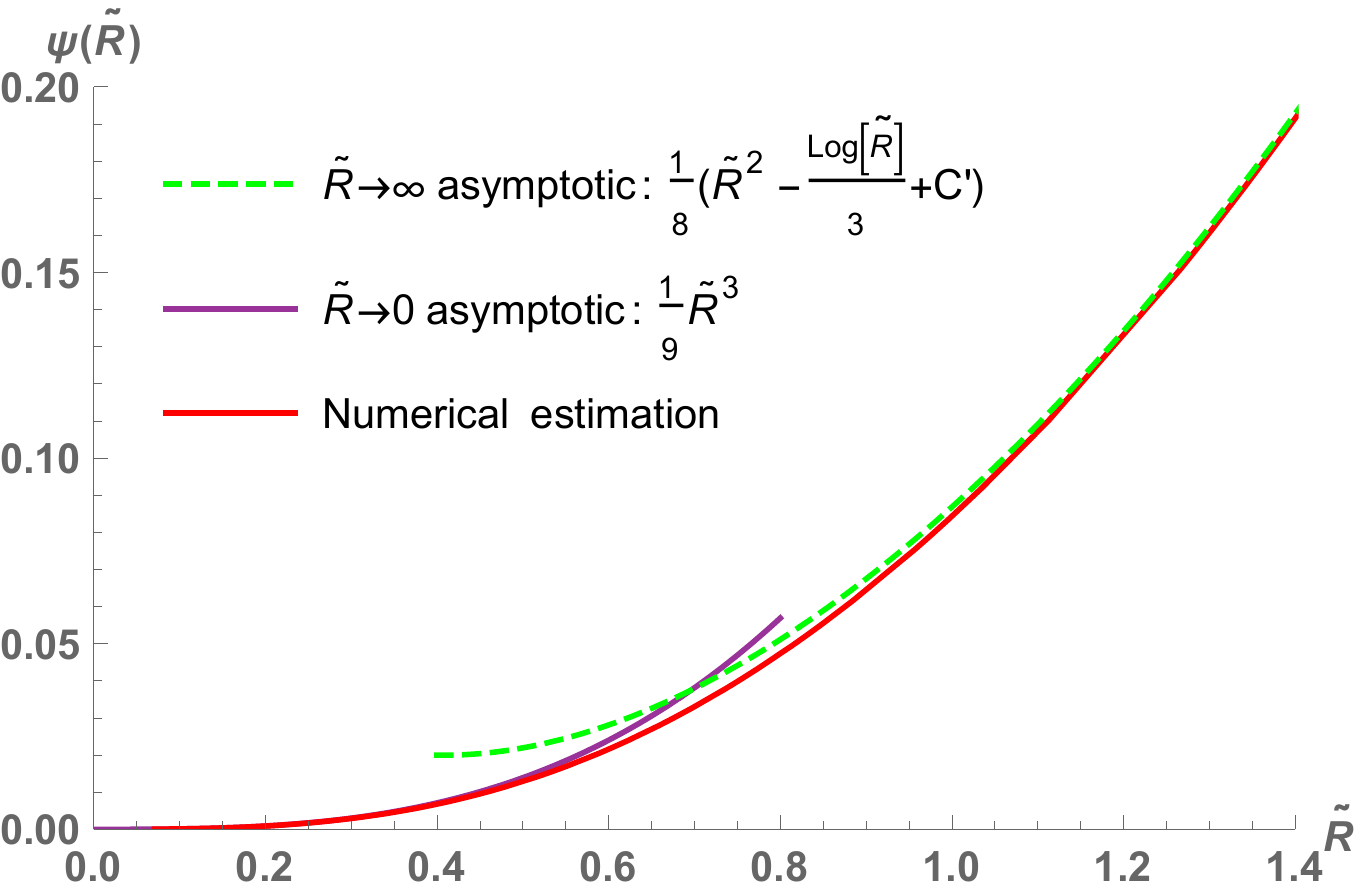}
\caption{Large deviation function $\Psi(\tilde{R})$ vs. $\tilde{R}=\frac{R}{\sqrt{2 \mu}}$ in dimension 2. The numerical evaluation (Red line) compares well with the small $\tilde{R}$ asymptotic (Purple line) itself matching with Eq. \eqref{largez}. It also fits well to the large $\tilde{R}$ behavior (Green line) from Eq. \eqref{LargeR}.}
\label{fig:largedev}
\end{figure}

\medskip

{\bf Small $\tilde R$ asymptotics.} Consider $\tilde R \ll 1$, in that case $v_{\max}(\tilde R) \simeq 2 \tilde R \ll 1$ hence $v \ll 1$ and $\alpha_v \simeq 2 v$.
In this limit we can thus change the integration variable in \eqref{result2} from $v$ to $\alpha=\alpha_v \simeq 2 v$ and obtain
\be \label{int3} 
\Psi(\tilde R) \simeq \int_0^{4 \tilde R}  d\alpha \frac{\alpha^{d-2}}{2^{2d} \Gamma(d-1)}
\Phi_+((\alpha^2( 4 (\frac{\tilde R}{\alpha})^2 - \frac{1}{4}) , \alpha)   
\ee 
Hence to study the small $\tilde R$ limit we need to consider the double limit of $\Phi_+(\alpha^2 y,\alpha)$ as 
$\alpha \to 0$ where $y=4 (\frac{\tilde R}{\alpha})^2 - \frac{1}{4}$ is fixed.
%We will now show that for $\alpha \to 0$, $\Phi_+(\alpha^2 y,\alpha) \simeq \frac{\alpha^2}{2} \phi_+(\gamma= \sqrt{1+ 4 y})$.
One has in this limit
\be  \label{comb} 
\Phi_+(\alpha^2 y,\alpha) = \frac{1}{2} (S((y+ \frac{1}{4}) \alpha^2,\alpha) - S(\frac{1}{4} \alpha^2,\alpha))
\ee 
Hence we need to calculate $S(z \alpha^2, \alpha)$ for $\alpha \ll 1$ and fixed $z$. One has 
\be \cos \theta \simeq \frac{3}{8} \sqrt{3} \alpha^2 \sqrt{z} %-\frac{9}{32} \alpha^3 \left(\sqrt{3} \sqrt{y}\right) + \dots \\
 \quad , \quad \theta \simeq \frac{\pi }{2}-\frac{3}{8} \alpha^2 \left(\sqrt{3} \sqrt{z}\right) %+\frac{9}{32} \sqrt{3} \alpha^3   \sqrt{y}+O\left(\alpha^4\right)
   \ee
which leads to 
   \be
U \simeq \frac{4}{3} (z \alpha^2 + 2 \alpha + 4) \cos^2(\frac{5 \pi}{6} - \frac{\alpha^2 \sqrt{3 z}}{8}) \simeq  4 + 2 \alpha + \alpha^2 ( - \sqrt{z} + z) + \dots 
\ee
%   Greg has
%\be
%U = 4 + 2 \alpha + \alpha^2 ( \frac{\pi \sqrt{3}}{8} - \sqrt{y} + y) + \dots 
%\ee
Inserting this expansion into the expression of $S$ given in Eq. (\ref{expr_S}) we obtain
\bea
S(z \alpha^2 , \alpha) = \frac{3}{2}+\alpha+\alpha^2 \left(\frac{\log (\alpha)}{2}+z-2
   \sqrt{z}+\frac{1}{4} \log (4
   z)\right)+O\left(\alpha^3\right) \;.
\eea 
From \eqref{comb} we now obtain 
\be 
\Phi_+(\alpha^2 y,\alpha) \simeq \frac{\alpha^2}{2} (y - \sqrt{1+ 4 y} + 1 + \frac{1}{4} \log(1+ 4 y) ) 
\ee 
which coincides exactly with $\frac{\alpha^2}{2} \phi_+(\gamma= \sqrt{1+ 4 y})$ where $\phi_+(\gamma)$ is given in
\eqref{phi_p}. Inserting this form in the integral \eqref{int3} we obtain that for small $\tilde R \ll 1$ 
\be
\Psi(\tilde R)  \simeq \frac{1}{2^{2 d+1} \Gamma(d-1)} \int_0^{4 \tilde R} d \alpha \alpha^d \phi_+(\gamma = \frac{4 \tilde R}{\alpha}) 
= \frac{2}{\Gamma(d-1)} \int_1^{+\infty} d \gamma \gamma^{-(d+2)} \phi_+(\gamma) \tilde R^{d+1} =
\kappa_d \tilde R^{d+1}
\ee
where $\kappa_d=2/((d+1)^2 \Gamma(d+1))$.
Hence we find $\Psi(\tilde R)  \simeq \kappa_d \tilde R^{d+1}$ at small
$\tilde R$ which matches exactly with the large radius limit 
of the microscopic regime in Eq.~\eqref{largez}.

\medskip

{\bf Large $\tilde R$ asymptotics.} We can also compute the large $\tilde{R}=\frac{R}{\sqrt{2\mu}}$ asymptotic of the large deviation function.

Let us rewrite \eqref{result2} in terms of the function $S(\zeta,\alpha)$ as
\bea \label{result3} 
&& \Psi(\tilde R) = \frac{1}{2}
\int_0^{v_{\rm max}(\tilde R)} dv\,\frac{v^{d-2} (1 - v)^2 }{2^{d+1} \Gamma(d-1)}  \left( S(\frac{4 \tilde R^2}{1-v},\alpha_v) - S((1 - \sqrt{1 + \alpha_v})^2,\alpha_v) \right) \\
&& \simeq \frac{1}{2^{d+2}\Gamma(d-1)}\int_{0}^{1}dvv^{d-2}(1-v)^2S(\frac{4\tilde{R}^2}{1-v},\alpha_v)+C_2
\eea 
where we used $\zeta_-=(1-\sqrt{1+\alpha_v})^2$ and $C_2$ is a $d$-dependent constant. In the second line we 
have considered $\tilde R>1/\sqrt{2}$. We recall that $\alpha_v = 2 v/(1-v)$. We computed $C_2$ numerically
%and found $C_2(d=2)=-0.085702$
%and $C_2(d=3)=-0.0159211$
and found $C_2(d=2)=-0.0232022$
and $C_2(d=3)=-0.0002961$.
 
The asymptotics of the functions $U(\zeta,\alpha)$ and $S(\zeta,\alpha)$ for large $\zeta$ and fixed $\alpha$ are found to be
\bea
&& U(\zeta,\alpha) %= \zeta + 2 (2 + \alpha) + O(\frac{\alpha^2}{\zeta}) \quad , \quad {\blue U(\zeta,\alpha) 
= \zeta + 4 + O(\frac{\alpha}{\zeta}) \\
&& S(\zeta,\alpha) = \zeta - \alpha \log \zeta + \frac{3}{2} + O(\frac{\alpha}{\zeta})
%\quad , \quad C_3(\alpha) = \frac{3+\alpha}{2} - \frac{\alpha^2}{8} + \log \frac{2}{2+\alpha} 
%\quad {\blue C_3(\alpha) = }
\eea
Inserting into \eqref{result3} one obtains 
\bea \label{result4} 
&& \Psi(\tilde R)  \simeq \frac{1}{2^{d+2}\Gamma(d-1)}
\int_{0}^{1}dv v^{d-2}(1-v)^2 \left(  \frac{4\tilde{R}^2}{1-v} - \alpha_v \log \frac{4\tilde{R}^2}{1-v} + \frac{3}{2} + O(\frac{\alpha_v (1-v)}{4\tilde{R}^2}) \right) + C_2
\eea 
We note that for each term the integral over $v$ is convergent. We can compute these integrals and obtain the large $\tilde R$ asymptotics as
\be \label{LargeR} 
\Psi(\tilde{R})=\frac{1}{2^d\Gamma(d+1)}\tilde{R}^2-\frac{(d-1)}{2^d\Gamma(d+2)}\log(\tilde{R}) + C + O(\frac{1}{\tilde R}^2) 
\ee
where the constant $C$ can be written as
\be
C = \frac{2^{-d-2} \left(-2 \left(d^2-1\right) H_d+d^2 (2-4 \log (2))+d+3+\log
   (16)\right)}{(d+1) \Gamma (d+2)} + C_2 %= \frac{C'}{2^{d+1}} \;,
\ee
where $H_d = \sum_{k=1}^d 1/k$ is the $d$-th harmonic number. 

%\be
%C = 
%\frac{2^{-d-3} (-d (d (d-6+\log (256))-5)-4 d (d+1) (\psi ^{(0)}(d)+\gamma
%   )+2+\log (256))}{(d+1) \Gamma (d+2)} + C_2 = \frac{C'}{2^{d+1}}
%\ee
%For $d=2$ we find $C=-0.0219367 + C_2(d=2)=-0.107639$ and for $d=3$ we find $C=-0.0117776 + C_2(d=3)=-0.0276987$.

For $d=2$ we find $C= -0.0149922 + C_2(d=2)=-0.0381944$ and for $d=3$ we find $C=-0.00895639 + C_2(d=3)=-0.00925249$.
%For $d=2$ we find $C=-0.0219367 + C_2(d=2)=-0.0451$ and for $d=3$ we find $C=-0.0117776 + C_2(d=3)=-0.01207$.
This asymptotics is compared to the exact result in Fig. \ref{fig:largedev}, with excellent agreement. 
%{\tiny $C'$ can be numerically estimated and is equal to $C'=-0.306$ in dimension 2}

The large deviation result \eqref{result2} is valid in the limit where $\mu$ (i.e., $N$) and $R$ are both large with $\tilde R = R/\sqrt{2 \mu}$ fixed. 
In the formula \eqref{LargeR} we have further taken the limit $\tilde R \to \infty$ in that large deviation regime. We will now study how it compares
with the calculation performed previously in a different regime, where $\mu$ (i.e., $N$) is fixed and $R$ is taken to infinity.

Consider the asymptotic behavior of the hole probability $P(R)$ in that regime, Eq.\eqref{harmoniclargeR}, which we rewrite as
\be \label{aa} 
P(R) \simeq C(\mu) \exp( - N R^2 + 2 M \log R)   
\ee 
We now consider the limit of large $\mu$ in that formula. In that limit, approximating the sums by integrals and
using the large $\ell$ asymptotics $g_d(\ell) \simeq \frac{2 \ell^{d-2}}{\Gamma(d-1)}$ and $m_\ell \simeq \frac{\mu - \ell}{2}$ one has
\bea
&& N = \sum_{\ell\ge 0}^{int(\mu-d/2)} g_d(\ell) m_{\ell} \simeq  \int_0^\mu d \ell \frac{2 \ell^{d-2}}{\Gamma(d-1)} \frac{\mu-\ell}{2} = 
\frac{\mu^d}{\Gamma(d+1)} \\
&& M=\sum_{\ell\ge 0}^{int(\mu-d/2)} g_d(\ell)(\ell+d/2-1) m_{\ell} \simeq  \int_0^\mu d \ell \frac{2 \ell^{d-1}}{\Gamma(d-1)} \frac{\mu-\ell}{2}
=
\frac{d-1}{\Gamma(d+2)}\mu^{d+1} 
\eea
Inserting into \eqref{aa} we obtain in terms of $\tilde R=R/\sqrt{2 \mu}$
\be \label{aa} 
P(R) \simeq C(\mu) \exp( - (2 \mu)^{d+1} \left(  \frac{1}{2^d\Gamma(d+1)} \tilde R^2 - \frac{d-1}{2^d\Gamma(d+2)} \log(\tilde R \sqrt{2 \mu}) 
\right) 
\ee 
This matches perfectly to the first two leading orders the formula \eqref{LargeR} obtained in the large deviation regime (the $\mu$ dependence
being cancelled by the constant $C(\mu)$) 
%{\red P: Gabriel see what you can say about the $\mu$ dependence and the constant. }

\subsection{Large deviation form in $d=1$}

In dimension $d=1$ done can compute the hole probability $P(R)$ that there are no fermions in the interval $[-R,R]$ for the harmonic potential
in the large deviation regime $R = O(\sqrt{\mu})$. We can exploit the relation between the positions of the fermions in the ground state and the eigenvalues of a GUE random matrix \cite{DeanReview2019}. 
We can then use the results of \cite{MMSV14} where the probability of the number of eigenvalues in an interval $[-L,L]$ was computed for
the GUE. Consider Eq. (56) in the Supp. Mat. of this work. We need to set $\kappa_L=0$, $a=0$, $b=\sqrt{2 + L^2}$, $\mu_2=\frac{1+ 2 L^2}{2}$,
$F(x)=x - \sqrt{ \frac{x^2 (x^2-b^2)}{x^2-L^2}} $, $\beta=2$. Performing the integral $\int_b^{+\infty} dx (F(x) - \frac{1}{x}) = \frac{1}{2} ( \log(2 + L^2) - 1 + \log 2)$,
setting $L=R/\sqrt{N}$ and $N \simeq \mu$ and 
and inserting one finds the remarkably simple result 
\be  \label{large_dev_d1}
P(R) \sim \exp \left(  - (2 \mu)^2 \Psi( \tilde R=\frac{R}{\sqrt{2 \mu}} ) \right) \quad , \quad \Psi(\tilde R) = \frac{1}{2} \tilde R^2 \;.
\ee 
It matches both the large $\tilde R$ behavior in \eqref{LargeR} setting $d=1$, and the small $\tilde R$ behavior 
$\Psi(\tilde R) \simeq \kappa_1 \tilde R^2$ with $\kappa_1=1/2$. Note that this large deviation function in (\ref{large_dev_d1}) was also obtained in Ref. \cite{MNSV2011} (see Eq. (132) of that paper).

\section{Hole probability in momentum space}

For experiments which use time of flight measurements (see e.g. \cite{flattrap}), it may be interesting to investigate the hole probability in momentum space. 

Consider the harmonic oscillator, with position variable ${\bf X}$ and momentum ${\bf P}$, where we have restored the dimensions. The single particle Hamiltonian is
\be
{\cal H} = \frac{{\bf P}^2}{2 m} + \frac{1}{2} m \omega^2 {\bf X}^2 = \hbar \omega H \quad , \quad H = \left( \frac{{\bf p}^2}{2} + \frac{{\bf x}^2}{2}  \right) 
\ee 
where we have defined the dimensionless variables $x,p$
\be 
{\bf x} = \alpha {\bf X} \quad, \quad {\bf p} = \frac{{\bf P}}{\hbar \alpha} \quad , \quad \alpha = \sqrt{ \frac{m \omega}{\hbar}} 
\ee 

We are interested here in the point process defined by the momenta ${\bf p}_i$, $i=1,\dots,N$ of $N$ noninteracting fermions in their ground state in 
this harmonic potential. One can define the hole probability 
$\tilde P(K ; {\bf p}_0)$ that the sphere in momentum space centered around some fixed momentum ${\bf p}={\bf p}_0$ and of radius $K$ is empty of
fermions. Since there is a perfect symmetry between ${\bf x}$ and ${\bf p}$ one obtains immediately that
\be \label{symmetry}
\tilde P(K; {\bf p}_0) = P(R=K ; {\bf x}_0 = {\bf p}_0)
\ee 
where $P(R;{\bf x}_0)$ is the hole probability for a sphere of radius $R$ in ${\bf x}$ space centered around ${\bf x}_0$. In the text
we have considered mainly the case ${\bf x}_0=0$, which we denoted $P(R)$. In the microscopic regime $R = O(1/k_F({\bf x}_0))$ we have further argued that $P(R;{\bf x}_0)$
takes the following scaling form in the large $N$ limit 
\be \label{scaling_x}
P(R;{\bf x}_0) \simeq {\sf P}_d( (a_d \rho({\bf x}_0))^{1/d} R)   \quad , \quad (a_d \rho({\bf x}_0))^{1/d} = k_F({\bf x}_0) = \sqrt{2 \mu - {\bf x}_0^2} \quad , \quad 
a_d = 2^d \pi^{d/2} \Gamma(1+ d/2)
\ee 
for any ${\bf x}_0$ in the bulk, i.e not in the region of the edges of the support of the density. Here $\rho({\bf x}_0)$ is the mean local fermion density in real space
at the point ${\bf x}_0$ and $\mu$ is the Fermi energy in units of $\hbar \omega$. This result implies, using the relation (\ref{symmetry}), that for "microscopic scales" in momentum space
\be \label{scaling_p}
\tilde P(K;{\bf p}_0) \simeq {\sf P}_d( (a_d \tilde \rho({\bf p}_0))^{1/d}  K) \quad , \quad (a_d \tilde \rho({\bf p}_0))^{1/d} = \sqrt{2 \mu - {\bf p}_0^2} 
\ee 
where the microscopic distance in momentum space is set by the inverse mean momentum density $K = O( (\tilde \rho({\bf p}_0))^{-1/d} )$ in the ground state. Note that
the density $\tilde \rho({\bf p})$ is defined as
\bea
\tilde \rho({\bf p}) = \frac{1}{(2 \pi)^d} \int d^{d} {\bf x} \, \theta(\mu - H({\bf x}, {\bf p})) 
\eea
and is also normalized to $N$ (as the density in real space). Note that the combinations of variables which enter inside the scaling functions 
in Eqs. (\ref{scaling_x}) and (\ref{scaling_p}) are dimensionless and thus invariant by a rescaling by $\alpha$, i.e these equations hold also
in terms of the original variables ${\bf X}$ and ${\bf P}$. The above results are exact for the harmonic potential, however in the microscopic regime we
expect that they hold more generally in the bulk of the Fermi gas for any smooth potential. 

\end{widetext}


\begin{thebibliography}{}
\bibitem{footnotetransition} A detailed calculation shows that the transition is of sixth order, i.e., the sixth
derivative is discontinuous.
\end{thebibliography}
\end{document}